\documentclass[journal]{IEEEtran}

\usepackage{amsmath,amssymb,amsfonts}
\usepackage{bbm}
\usepackage[hidelinks,colorlinks=true,linkcolor=blue,citecolor=blue,urlcolor=black]{hyperref}
\usepackage[nolist]{acronym}
\usepackage{graphicx}
\usepackage{amsmath}
\usepackage{amssymb}
\usepackage{textcomp}
\usepackage{comment}
\usepackage{subfigure}
\usepackage{subfiles} 
\usepackage{xcolor}
\usepackage{textgreek}

\def\BibTeX{{\rm B\kern-.05em{\sc i\kern-.025em b}\kern-.08em
    T\kern-.1667em\lower.7ex\hbox{E}\kern-.125emX}}
\usepackage[linesnumbered,ruled,vlined]{algorithm2e}
\usepackage{listings}
\usepackage[noend]{algpseudocode}
\SetCommentSty{mycommfont}
\SetKwInput{KwInput}{Input}                % Set the Input
\SetKwInput{KwOutput}{Output}              % hset the Output
\usepackage{subcaption}

\usepackage{balance}
\usepackage{cite}
\usepackage{soul}
\usepackage{multirow} % for multi row in tables
\usepackage[T1]{fontenc} % for << symbol
\usepackage{cuted}  % for strip comment related to equation
\usepackage{relsize}

\usepackage{booktabs}
\usepackage{longtable}
\usepackage{acronym}

\input{Acro}
\acresetall
\begin{document}

\title{A High Altitude Platform-Based 3D Geometrical Channel Model for Beamforming Characterization in Future 6G Flying Ad-Hoc Networks}
\author{
    \IEEEauthorblockN{Muhammet Kırık}
    \href{https://orcid.org/0000-0003-2431-8075}{\includegraphics[scale=0.01]{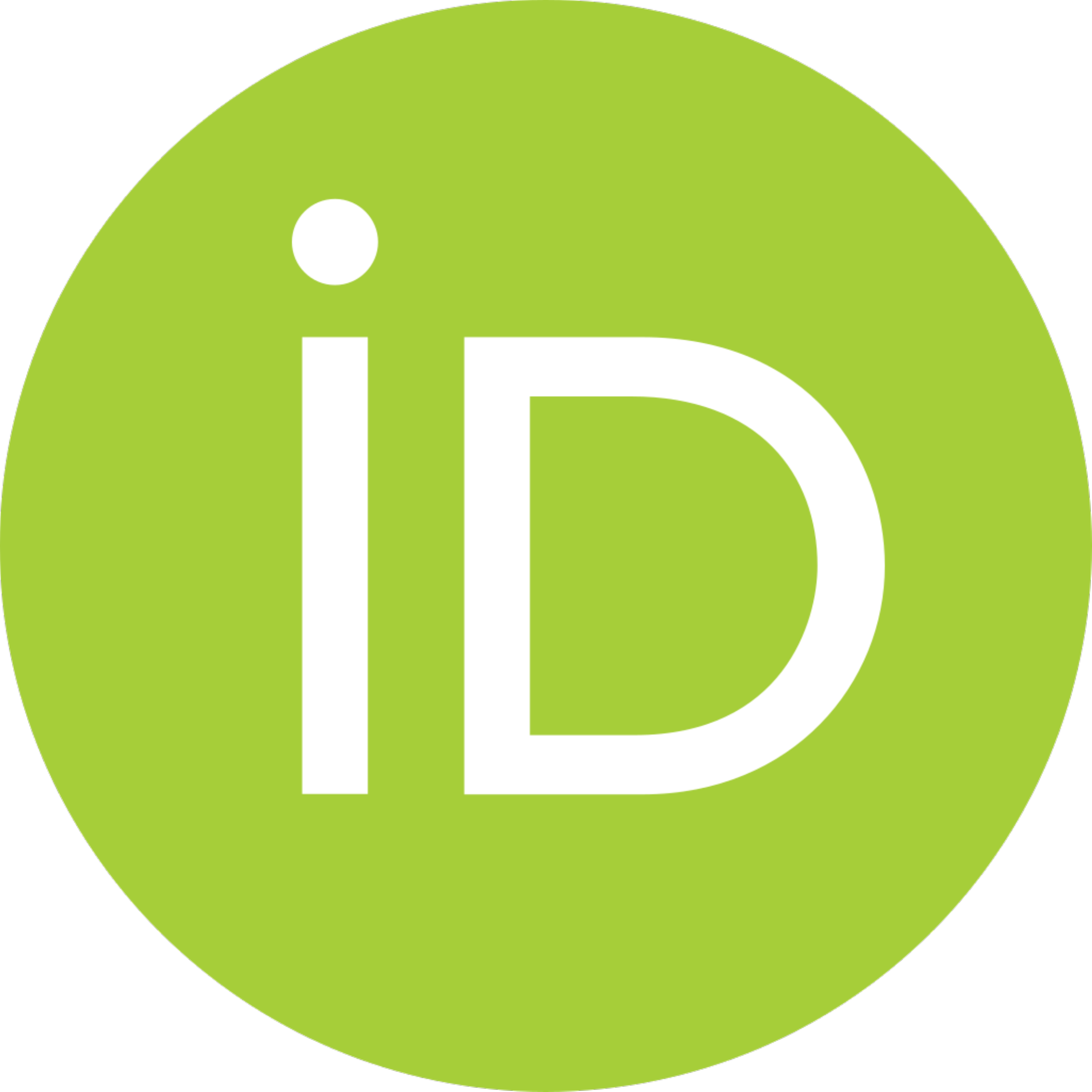}},
    \IEEEauthorblockN{Liza Afeef}
    \href{https://orcid.org/0000-0001-7679-0326}{\includegraphics[scale=0.01]{1024px-ORCID_iD.svg.pdf}},
	\IEEEauthorblockN{and}
	\IEEEauthorblockN{H\"{u}seyin Arslan}
    \href{https://orcid.org/0000-0001-9474-7372}{\includegraphics[scale=0.01]{1024px-ORCID_iD.svg.pdf}},
	\IEEEmembership{Fellow, IEEE}

\thanks{ The authors are with the Department of Electrical and Electronics Engineering, Istanbul Medipol University, 34810 Istanbul, Turkey (e-mail:
muhammet.kirik@medipol.edu.tr; liza.shehab@std.medipol.edu.tr; huseyinarslan@medipol.edu.tr).}
}

\maketitle

\begin {abstract} Growing requirements of future wireless communication systems, such as high data rates, high reliability, and low latency, make the active usage of \ac{NTN} an inevitable necessity. In this regard, \acp{HAP} have drawn great attention in recent years due to their unique characteristics such as high coverage, long operational durability, and ad-hoc movement. However, for the active usage of \acp{HAP}, channel models for their various usage scenarios must be well-defined, especially in those cases where the sophisticated \ac{MIMO} techniques, such as beamforming, are utilized to increase the data rate. Therefore, in this study, an \ac{A2A} \ac{3D} geometrical channel model is proposed to characterize the beamforming capabilities of non-stationary \ac{HAP} networks operating at \ac{mmWave} frequency band. In this regard, the \ac{3D} geometry of the two \acp{HAP} in the air is analyzed, and the effect of Doppler due to the movement of \acp{HAP} is interrogated as well as its effect on the \ac{SNR}. The final outputs of this study show that the proposed \ac{A2A} channel model is applicable to characterize the future \ac{6G} \ac{HAP} networks when the \ac{mmWave} is used to utilize beamforming with a large number of antennas. 
\end {abstract}


\IEEEpeerreviewmaketitle
\setcounter{section}{0}
\section {Introduction} \label{sec:introduction}

\IEEEPARstart{D}{ue} to the uneven distribution of \ac{IoT} devices and internet users in various environments, such as urban centers, hilly terrains, rural villages, vast oceans, and many other challenging environments, ensuring adequate coverage becomes a crucial factor in providing high \ac{QoS} \cite{li2018uav}. In this regard, \ac{5G} has introduced the usage of \ac{NTN} for the first time and identified the characteristics of \ac{NR} in \ac{NTN} to eradicate the possible challenges on the way to provide full coverage \cite{kirik2023inter,demir2022performance}. However, even though there are some early initiatives to implement \ac{NTN} into existing communication networks, it does not appear like the current \ac{5G} standards that are mainly composed of terrestrial \acp{BS} will provide a satisfactory solution to cover the entire surface of the Earth in the \ac{6G} era \cite{zhang20196g}.

On the other hand, the adoption of \ac{mmWave} and \ac{MIMO} systems in \ac{5G} and \ac{6G} networks is motivated by the growing demand for wireless communication and the scarcity of available spectrum caused by the proliferation of connected devices and Internet applications \cite{afeef2022beam}. With the introduction of \ac{5G}, \ac{mmWave}, operating in the 30-300 GHz range, is harnessed to provide abundant bandwidth, low latency, and gigabit-speed connectivity for billions of devices \cite{wang2018millimeter}. However, exploiting the \ac{mmWave} spectrum presents challenges such as pathloss, penetration loss, shadowing, atmospheric attenuation, and hardware limitations. In response, both industry and academia have worked on innovative \ac{MIMO} techniques and channel models to enable the use of \ac{mmWave} in outdoor \ac{5G} cellular networks.

% now enlerge the MIMO in HAP systems
Implementing \ac{mmWave} and \ac{MIMO} technologies on \ac{HAP} communication systems can improve the spectral efficiency. However, within the advent of the \ac{6G} era, the requirement for novel transmission techniques and channel models for long-distance non-terrestrial \acf{A2A} and \ac{A2G} applications, is more than ever to complete the vision of the \ac{5G} and prepare the world for the next generation communication networks.

% However, most of these efforts were based on terrestrial base stations, and, as of today's conditions, their usage is still very limited. Therefore, within the advent of the \ac{6G} era, the requirement for novel transmission techniques and channel models for \ac{mmWave}, not only for indoor or small cell terrestrial based outdoor applications but also for long distance non-terrestrial \acf{A2A} and \ac{A2G} applications, is more than ever to complete the vision of the \ac{5G} and prepare the world for the next generation communication networks.\par

In this regard, this paper analyses and models the beamforming usage at \ac{mmWave} band in a scenario where two \acp{HAP} communicate through an \ac{A2A} channel. It should be noted that the majority of \ac{A2A}/\ac{A2G} channel models proposed in the literature are developed for low altitude \acp{UAV}. However, \acp{HAP} have their unique advantages that make them more preferable than both spaceborne terminals such as \ac{LEO} satellites or low altitude airborne \acp{UAV} such as drones. Therefore, the proposed channel model is considered on the basis of \acp{HAP}' communication.

\subsection{Literature Review}
%Within the upcoming \ac{6G} era, the interest in the \ac{UAV}-based communication has drastically increased. In this manner, many different designs, algorithms, and channel models for both \ac{A2G} and \ac{A2A} communication have been proposed in the literature to prepare the communication society for the active usage of flying access points in the future.
In this subsection, some of the related state-of-the-art works in the literature are reviewed with their potential and limitations.

% To begin with, the series of papers by \textit{Ma, et al.} given in \cite{ma2019wideband,ma2020three,ma2021non} can be investigated, since the works presented in these papers are the most relevant to the proposed channel model.

In \cite{ma2019wideband}, a geometrical \ac{A2A} channel model is presented for \ac{UAV} communication. Although the proposed channel model provides some enlightening insights about non-stationary \ac{UAV} networks, this model misses the opportunity to investigate the characteristics of \ac{mmWave} band. Differing from \cite{ma2019wideband}, the work in \cite{ma2020three} and \cite{ma2021non} represents a geometrical \ac{A2A} channel model for non-stationary \ac{UAV} networks that utilize mmWave \ac{MIMO} technology. Even though the proposed channel model in these works provides some valuable insights about the utilization of \ac{mmWave} band with \ac{MIMO} in non-stationary \ac{UAV} communications, the characterization of beamforming, which is one of the key techniques for the usage of higher frequency bands, is not conducted by the authors.

% Another work that provides an elaborate channel model analysis for \ac{UAV} networks is given by \textit{Dabiri et al.} in \cite{dabiri2020analytical}. The work provides a geometrical channel model where three scenarios are mainly concerned, which are an \ac{A2A} link scenario, a ground-to-air-to-ground link scenario, and a scenario where the communication between two \acp{UAV} is amplified by a \ac{UAV} that serves as a relay in the middle. The proposed channel models for these scenarios are conducted under the assumption that the \acp{UAV} are capable of performing directional beamforming by deploying multiple antenna elements. However, even though the work of \textit{Dabiri et al.} in \cite{dabiri2020analytical} extensively shows the channel characteristics of \ac{UAV} networks from the beamforming perspective, due to the fact that the paper ignores the mobility of the \acp{UAV} and focuses on a hovering scenario, it leaves some unlit areas, which can be observed only in mobile scenarios such as Doppler shift, beam shift, etc. 

\textit{Dabiri et al.} in \cite{dabiri2020analytical} provided an elaborate channel model analysis for \ac{UAV} networks using a geometrical approach. The work focuses on three scenarios: an \ac{A2A} link scenario, a ground-to-air-to-ground link scenario, and a scenario where the communication between two \acp{UAV} is amplified by a \ac{UAV} that serves as a relay in the middle. The proposed channel models for these scenarios are conducted under the assumption that the \acp{UAV} are capable of performing directional beamforming by deploying multiple antenna elements. However, the model ignores the mobility of the \acp{UAV} and focuses on a hovering scenario which leaves some unlit areas that can be observed only in mobile scenarios such as Doppler shift, beam shift, etc. \par
Moreover, some channel models for low-altitude \ac{A2A} and \ac{A2G} communication links are presented in \cite{jiang2019three,zeng20173d,mao20213d, jiang2023physics,cao2021non} along with some channel models for \ac{HAP} networks given in \cite{lian2021non, lian2016novel, lian2018non}. Although the existing \ac{A2A} literature is heavily based on low-altitude airborne terminals, which provides some insight into the channel characteristics of inter-\ac{HAP} communication links they cannot be adopted directly to characterize the high-altitude \ac{A2A} channel for \acp{HAP} networks.\par

\subsection{Contributions}

The contributions of this paper can be listed as follows:
\begin{itemize}
     \item [i.)] At higher frequency bands such as \ac{mmWave}, characteristics of the channel heavily depend on environmental factors such as scatterers and reflectors. This is due to the fact that the short wavelength at these bands causes the signal to act as a beam and deviate from its direction rather than creating a multipath. To characterize these deviations, geometrical channel models are heavily used in the literature rather than deterministic or stochastic channel models. As stated earlier, the characteristics of low altitude \acp{UAV} and \acp{HAP} cannot be evaluated by the same channel model due to their environmentally identifying differences. Due to the fact that at the altitude that \acp{HAP} operate the possible scatterers and reflectors such as constructions, vehicles, clouds, etc. do not exist, the geometry of the transmitted signal beam is required to be specifically analysed for \acp{HAP}. Therefore, this paper provides an elaborate \ac{3D} geometrical channel model analysis for those scenarios where two non-stationary \acp{HAP} communicate through an \ac{A2A} channel at \ac{mmWave} band and the beamforming is performed on the transmitted signal; 
    \item [ii.)]  A comprehensive mobility model is proposed for \acp{HAP} that takes into account the altitude, speed, and trajectory. The model allows the characterization of the unrestricted movement capabilities of \acp{HAP}, which is essential for understanding their behavior in the presence of various environmental factors. This contribution provides a tool for predicting and optimizing the performance of \acp{HAP} based on their movement characteristics;
    \item [iii.)] The paper mathematically analyzes \ac{MIMO} and beamforming characteristics between a transmitter and receiver \ac{HAP}, and consolidates the results with simulations. Specifically, the authors investigate the impact of the relative positions, orientations, and beamforming parameters of the two \acp{HAP} on the quality of the communication link. Beamforming is a critical technique for enhancing the performance of wireless communication systems, and the analysis conducted in this paper provides insight into how it can be used to optimize \ac{HAP} networks;
    \item [iv.)] The paper investigates the effect of Doppler on the performance of the received signal when the beam direction is misaligned. The authors consider both the angular and frequency components of the Doppler effect and examine how they affect the quality of the received signal. Doppler can cause significant degradation in the performance of wireless communication systems, and the analysis in this paper provides a better understanding of how it impacts \ac{HAP} networks. The results of this analysis can be used to inform the design of more robust communication systems that can mitigate the effects of Doppler;
    %\item [v.)] Overall, this paper provides an elaborate   \ac{3D} geometrical channel model analysis for those scenarios where two non-stationary \acp{HAP} communicate through an \ac{A2A} channel at \ac{mmWave} band and the beamforming is performed on the transmitted signal. From this aspect, this is the first paper in the literature that provides a beamforming analysis at \ac{mmWave} band for future \acp{FANET} deploying \acp{HAP}.
\end{itemize} 
\par
\subsection{Organization}
The rest of this paper is organized as follows. In Section \ref{Sec:systemModel}, the system model of the proposed channel model is given by explaining the operation scenario in detail, providing a mobility model, and interrogating the effect of Doppler due to mobility. In Section \ref{Sec:Analysis}, the performance metrics of the proposed channel model are mathematically analyzed. In Section \ref{Sec:pdf}, the mathematical analyses are continued by deriving the probabilistic characteristics of the proposed channel model. In Section \ref{Sec:simulation}, the simulation results of the proposed channel model are shown along with a discussion. Lastly, in Section \ref{Sec:conclusion}, the paper is concluded.\footnote{\textit{Notations}: Note that vectors are denoted by bold-small letters, matrices are denoted by bold-capital letters, scalar numbers are denoted by non-bold capital letters, and complex numbers or elements indexing are denoted by non-bold small letters based on the context. Also, $H$ is given to denote the Hermitian transpose a corresponding matrix, whereas $\otimes$ is the Kronecker product between two matrices, and $\text{tr}(\cdot)$ is the trace of a matrix.} 

\section{System Model} 
\label{Sec:systemModel}

\subsection{Transceiver Model}
In this scenario, it is considered that the two highly mobile \acp{HAP} are communicating to each other through an A2A channel with altitudes $\mathbbm{h}_1$ and $\mathbbm{h}_2$  respectively. Due to the fact that the nature of \acp{HAP} allows them to move without any restricted trajectory, for each time iteration ($t_i$), positions of the \acp{HAP} are assumed to be unpredictable. In this scenario, due to the severe pathloss that is caused by the short wavelength of the transmitted signal, the \ac{3D} distance ($D_{3D}$) between the two \acp{HAP} is assumed to be limited and the \ac{2D} distance ($D_{2D}$) between them is assumed to chance by depending on $D_{3D}$ along with $\mathbbm{h}_1$ and $\mathbbm{h}_2$.\par

The main purpose of this study is to characterize the A2A channel for beamforming applications in \acp{FANET}. Therefore, the carrier frequency ($f_c$) is selected as a \ac{mmWave} band frequency and each \ac{HAP} is assumed to be equipped with $N\times N$ directional \ac{URPA} antennas. Due to the fact that the operating frequency is at mmWave band and the beamforming is applied on the transmitted signal, the position vectors of the two \acp{HAP} in each time iteration must be accurately identified.\par

For this reason, a 3D approach is adopted in the proposed channel model and the time varying direction angles of \ac{HAP}-1 and \ac{HAP}-2 are assumed to be dependent on both azimuth and elevation domains. The direction angles of \ac{HAP}-1 in azimuth and elevation domains are given as $\mathcal{D}_{A_{1}}$ and $\mathcal{D}_{E_{1}}$ respectively, whereas these angles for \ac{HAP}-2 are provided as $\mathcal{D}_{A_{2}}$ and $\mathcal{D}_{E_{2}}$. \par

Since the movement directions of the \acp{HAP} in time are random and independent, the transmitted signal beams from one \ac{HAP} will not always be guaranteed to be perfectly aligned with the receiver of the other \ac{HAP}. This is due to the fact that if the antenna directions are considered to be the same as the moving directions of the \acp{HAP}, the communication between them will be corrupted for every change in direction. Therefore, in order to provide the continuity of communication between the two \acp{HAP}, the steering directions of the antennas located in the \acp{HAP} are required to be adjusted by manipulating the phase angles of the antennas in accordance with the direction angles of the \acp{HAP}. In this regard, the angles $(\theta_{1},\phi_{1})$ are given as the phase angles of \ac{HAP}-1 in azimuth and elevation domains and  $(\theta_{2},\phi_{2})$ are given as the phase angles of \ac{HAP}-2 in azimuth and elevation domains respectively.\footnote{In Section \ref{Sec:Analysis}, \ac{HAP}-1 is considered as the transmitter and \ac{HAP}-2 is considered as the receiver. Therefore, to preserve the flow of the paper and keep the readers understanding high, the notations of $(\theta_{1},\phi_{1})$ and $(\theta_{2},\phi_{2})$ are changed accordingly in Section \ref{Sec:Analysis} and renotated as $(\theta_{1}=\theta^{Tx},\phi_{1}=\phi^{Tx})$ and $(\theta_{2}=\theta^{Rx},\phi_{2}=\phi^{Rx})$.} \par

\begin{comment}
Note that this paper provides a full system analysis for the characterization of beamforming in \acp{HAP} rather than providing a solution to eliminate the degrading effects of beam misalignment. Therefore, in the following analysis, the alignment of the transmitted beam is assumed to be both achieved and failed and the resulting outcomes are interpreted accordingly. 
\end{comment}
\vspace{-0.1cm}
\subsection{Mobility Model}
\label{Mobility Model}
In this study, the Gauss-Markov mobility model \cite{liang1999predictive} is exploited in conjunction with a 3D random walk mobility model \cite{chiang20042}. The reason why such a collaborative mobility model is adopted in this study is that even though the Gauss-Markov mobility model is widely used in the literature to demonstrate the temporal relation of the moving objects, in real-life scenarios, the probability of unexpected sudden rotations during the movement of these objects is substantially high and needs to be taken under consideration. Therefore, in this study, a random walk mobility model where the next maneuver of the object varies independently than the previous action is exploited in a harmony with the Gauss-Markov mobility model. \par
In this regard, for both \acp{HAP}, the velocity, azimuth direction, and elevation direction vectors are calculated in each time iteration, $t_i$, with respect to the values of these vectors at time instant $t_{i-1}$ as follows \cite{camp2002survey}

\begin{equation}
\begin{aligned}
\label{velocityModel}
  &\mathcal{V}_{(1,2)}(t_i)=\alpha_{\mathcal{V}_{(1,2)}}\mathcal{V}_{(1,2)}(t_i-1)+\\&
  \left(1-\alpha_{\mathcal{V}_{(1,2)}}\right)\mu_{\mathcal{V}_{(1,2)}}+\sqrt{1-\left(\alpha_{\mathcal{V}_{(1,2)}}\right)^2}X_{(1,2)_{i-1}},
\end{aligned}
\end{equation}

\begin{equation}
\begin{aligned}
\label{azimuthDirectionModel}
  &\mathcal{D}_{A_{(1,2)}}(t_i)=\alpha_{\mathcal{D}_{A_{(1,2)}}}\mathcal{D}_{A_{(1,2)}}(t_i-1)+\\&
  \left(1-\alpha_{\mathcal{D}_{A_{(1,2)}}}\right)\mu_{\mathcal{D}_{A_{(1,2)}}}+\sqrt{1-\left(\alpha_{\mathcal{D}_{A_{(1,2)}}}\right)^2}Y_{(1,2)_{i-1}},
\end{aligned}
\end{equation}

\begin{equation}
\begin{aligned}
\label{elevationDirectionModel}
  &\mathcal{D}_{E_{(1,2)}}(t_i)=\alpha_{\mathcal{D}_{E_{(1,2)}}}\mathcal{D}_{E_{(1,2)}}(t_i-1)+ \\&
  \left(1-\alpha_{\mathcal{D}_{E_{(1,2)}}}\right)\mu_{\mathcal{D}_{E_{(1,2)}}}+\sqrt{1-\left(\alpha_{\mathcal{D}_{E_{(1,2)}}}\right)^2}Z_{(1,2)_{i-1}},
\end{aligned}
\end{equation}
where $i\in \{1,2,3,...\}$ is the iteration factor that is used to denote each time iteration, $t_i$. While $\mathcal{V}_{(1,2)}(t_i)$, $\mathcal{D}_{A_{(1,2)}}(t_i)$, and $\mathcal{D}_{E_{(1,2)}}(t_i)$ are the velocity, azimuth direction, and elevation direction vectors at time instant $t_i$ for both \ac{HAP}-1 and \ac{HAP}-2 respectively, $\mathcal{V}_{(1,2)}(t_{i-1})$, $\mathcal{D}_{A_{(1,2)}}(t_{i-1})$. $\mathcal{D}_{E_{(1,2)}}(t_{i-1})$ represent the values of these vectors at time instant $t_{i-1}$. $\mu_{\mathcal{V}_{(1,2)}}$, $\mu_{\mathcal{D}_{A_{(1,2)}}}$, and $\mu_{\mathcal{D}_{E_{(1,2)}}}$ are given as the asymptotic mean values of velocity, azimuth direction deviation, and elevation direction deviation respectively for both \ac{HAP}-1 and \ac{HAP}-2 when $i$ approaches to infinity. The parameters $X,Y,Z$ are given as the Gaussian distributed random variables with zero mean and unity variance, whereas the parameters of $\alpha$, namely $\alpha_{\mathcal{V}_{(1,2)}}$,  $\alpha_{\mathcal{D}_{A_{(1,2)}}}$,  $\alpha_{\mathcal{D}_{E_{(1,2)}}}$ are the tuning parameters of \ac{HAP}-1 and \ac{HAP}-2 that have a range between $[0, 1]$, which defines the level of randomness and memory \cite{ariyakhajorn2006comparative}.\par 
%To clarify, when the value of $\alpha$ is closer to zero, the next movement of the \acp{HAP} becomes more challenging to predict, which means that the model becomes memoryless. In these cases, the values of velocity and direction vectors are dependent on the average values of their vectors and the Gaussian distributed parameters. However, if the value of $\alpha$ is closer to one, prediction of the next movement becomes easier, which means the model has a high memory that is inherited from the previous movement. In these cases, the new velocity and direction values at time instant $t_{i}$ are either identical as their corresponding values at previous time instant $t_{i-1}$, or extremely close to the previous velocity and direction vectors' values. \par

By assigning the initial values of velocity and direction vectors, the formulations of the Gauss-Markov mobility model given in (\ref{velocityModel}), (\ref{azimuthDirectionModel}), (\ref{elevationDirectionModel}) can be rewritten as \cite{ma2019wideband}

\begin{equation}
\begin{aligned}
\label{velocityModel2}
  &\mathcal{V}_{(1,2)}(t_i)=(\alpha_{\mathcal{V}_{(1,2)}})^i\mathcal{V}_{(1,2)}(t_0)+\\&
  \biggl(1-(\alpha_{\mathcal{V}_{(1,2)}})^i\biggl) \mu_{\mathcal{V}_{(1,2)}}+\sqrt{1-\left(\alpha_{\mathcal{V}_{(1,2)}}\right)^2}\\&
  \sum_{j=0}^{i-1}(\alpha_{\mathcal{V}})^{i-j-1}X_j,
\end{aligned}
\end{equation}

\begin{equation}
\begin{aligned}
\label{azimuthDirectionModel2}
  &\mathcal{D}_{A_{(1,2)}}(t_i)=(\alpha_{\mathcal{D}_{A_{(1,2)}}})^i\mathcal{D}_{A_{(1,2)}}(t_0)+\\&
  \biggl(1-(\alpha_{\mathcal{D}_{A_{(1,2)}}})^i\biggl) \mu_{\mathcal{D}_{A_{(1,2)}}}+\sqrt{1-\left(\alpha_{\mathcal{D}_{A_{(1,2)}}}\right)^2}\\&
  \sum_{j=0}^{i-1}(\alpha_{\mathcal{D}_{A_{(1,2)}}})^{i-j-1}Y_j,
\end{aligned}
\end{equation}

\begin{equation}
\begin{aligned}
\label{elevationDirectionModel2}
  &\mathcal{D}_{E_{(1,2)}}(t_i)=(\alpha_{\mathcal{D}_{E_{(1,2)}}})^i\mathcal{D}_{E_{(1,2)}}(t_0)+\\&
  \biggl(1-(\alpha_{\mathcal{D}_{E_{(1,2)}}})^i\biggl) \mu_{\mathcal{D}_{E_{(1,2)}}}+\sqrt{1-\left(\alpha_{\mathcal{D}_{E_{(1,2)}}}\right)^2}\\&
  \sum_{j=0}^{i-1}(\alpha_{\mathcal{D}_{E_{(1,2)}}})^{i-j-1}Z_j,
\end{aligned}
\end{equation}
%where $\mathcal{V}_{(1,2)}(t_i)$, $\mathcal{D}_{A_{(1,2)}}(t_i)$, and $\mathcal{D}_{E_{(1,2)}}(t_i)$ are the initial velocity, azimuth direction, and elevation direction values of \ac{HAP}-1 and \ac{HAP}-2 respectively at time instant $t_0$.
\par

In this study, it is assumed that the movement paths of the \acp{HAP} are predefined for a random scenario. Therefore it is expected that there are some sudden rotations during the movement of the \acp{HAP}. In order to characterize these sudden rotations, the random walk mobility model is exploited in conjunction with the Gauss-Markov mobility model. Even though the number of sudden rotations and the duration of the predictable motion of the \acp{HAP} are randomly selected, it should be noted that the randomness that is mentioned in this mobility model is predefined for the general trajectory. This means that once the trajectory is defined by authorities for a specific scenario, the movement path is restricted to follow the predefined trajectory. However, unexpected changes in the environment and the mission updates may still force the \acp{HAP} to deviate their directions from the predefined trajectory \cite{bekmezci2013flying}. The clarification of this trajectory definition and the randomness mentioned in this mobility model is crucial due to the fact that allowing \acp{HAP} to fly in the air randomly may cause some safety issues, which is not the case proposed in this paper.  \par

\begin{figure*}[h]
    \centering
    \subfigure[Velocity vector]{{\includegraphics[scale=0.18]{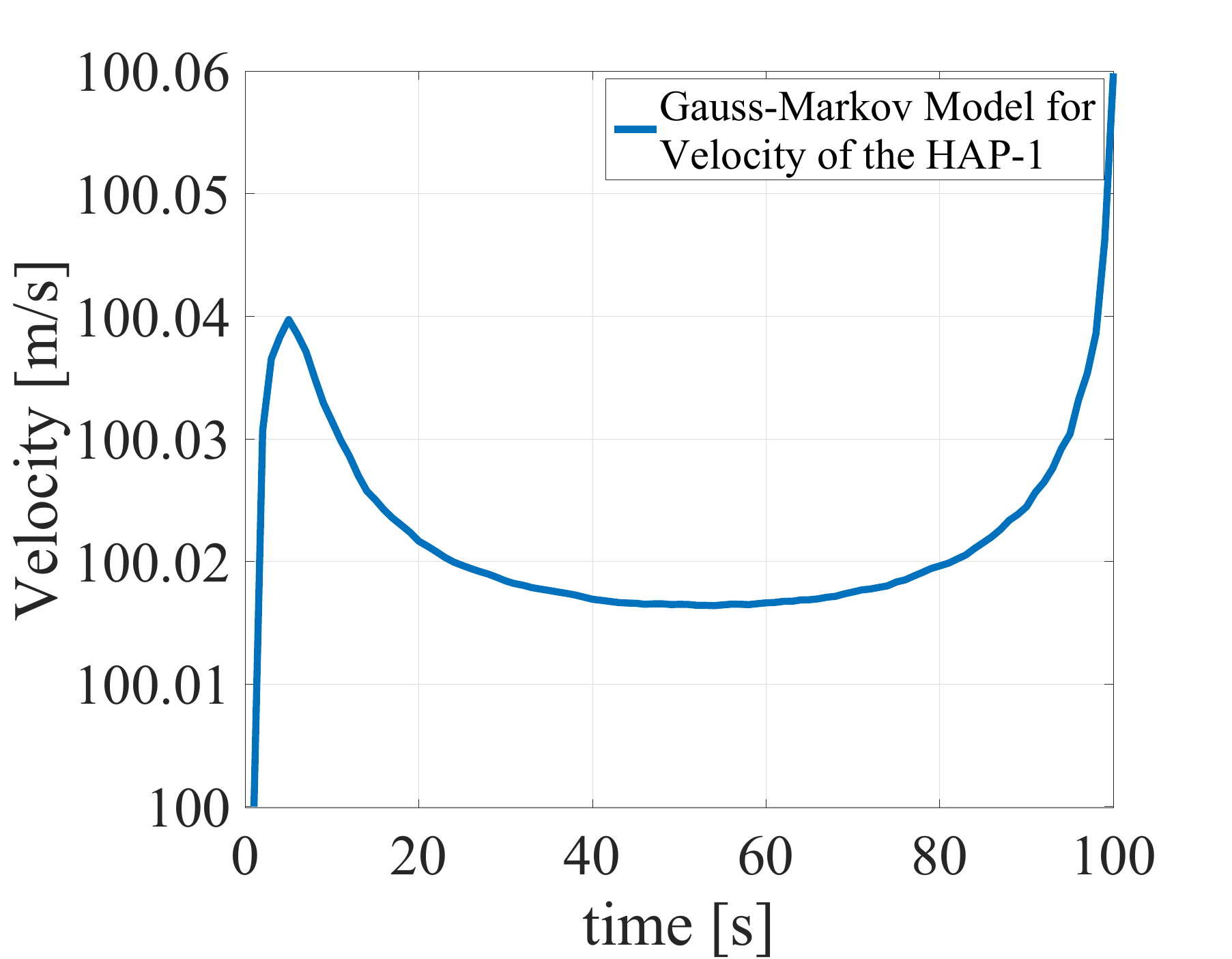}}}%
    \qquad
    \subfigure[Direction vector in azimuth domain]{{\includegraphics[scale=0.18]{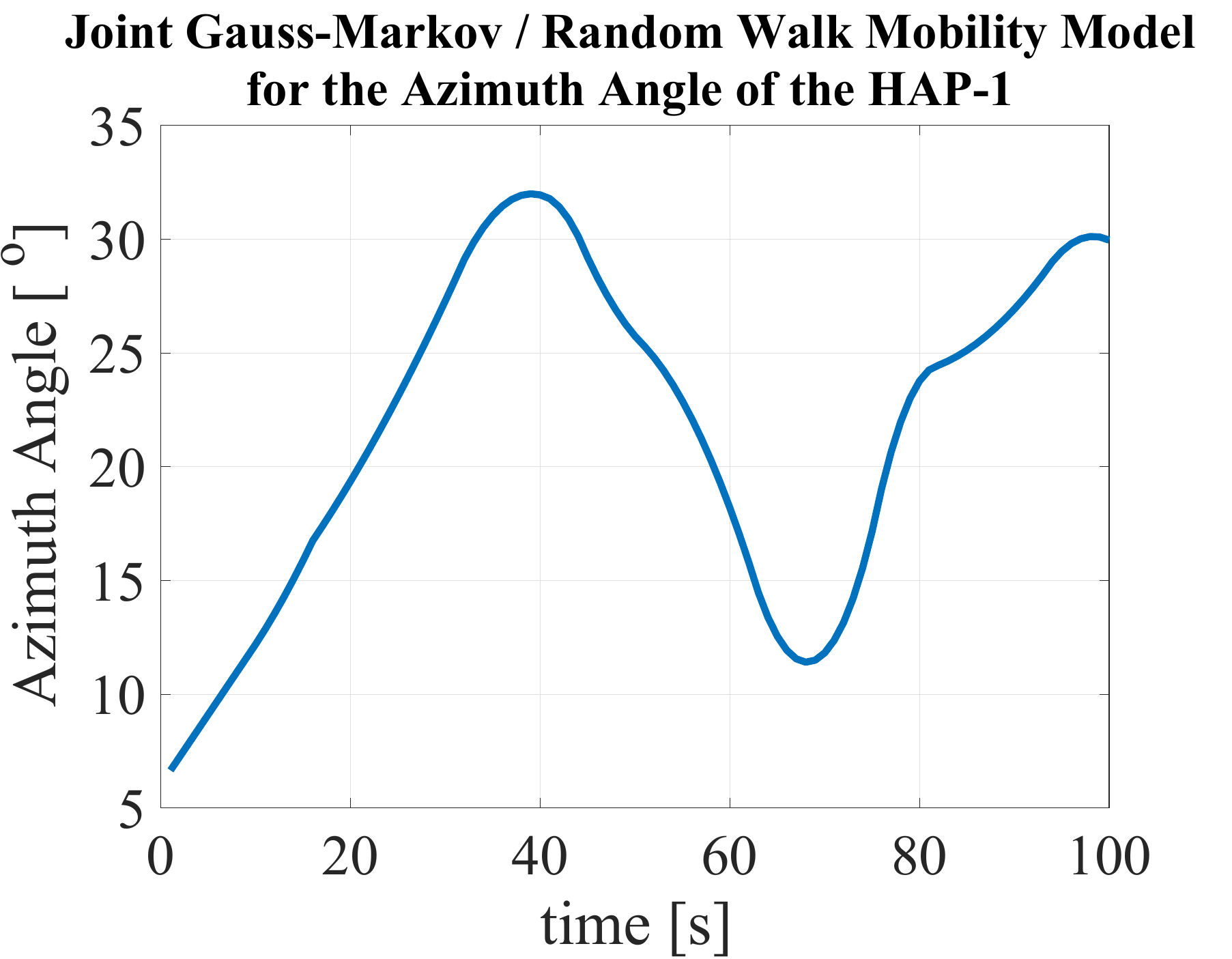}}}%
    \qquad
    \subfigure[Direction vector in elevation domain]{{\includegraphics[scale=0.18]{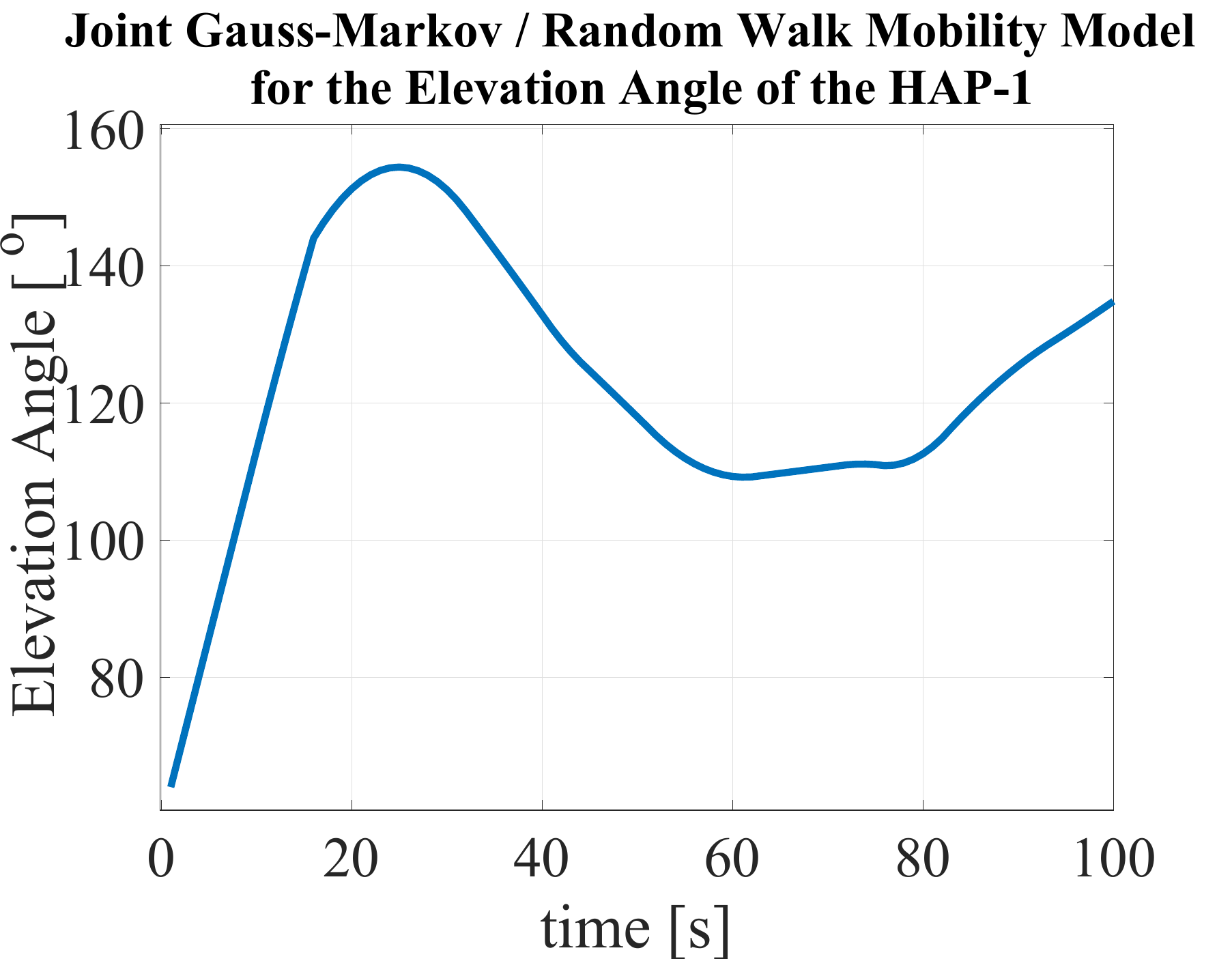}}}%
    \qquad
    \caption{Position vectors of \ac{HAP}-1}%
    \label{vectorsHAP1}%
\end{figure*}

\begin{figure*}[h]
    \centering
    \subfigure[Velocity vector]{{\includegraphics[scale=0.18]{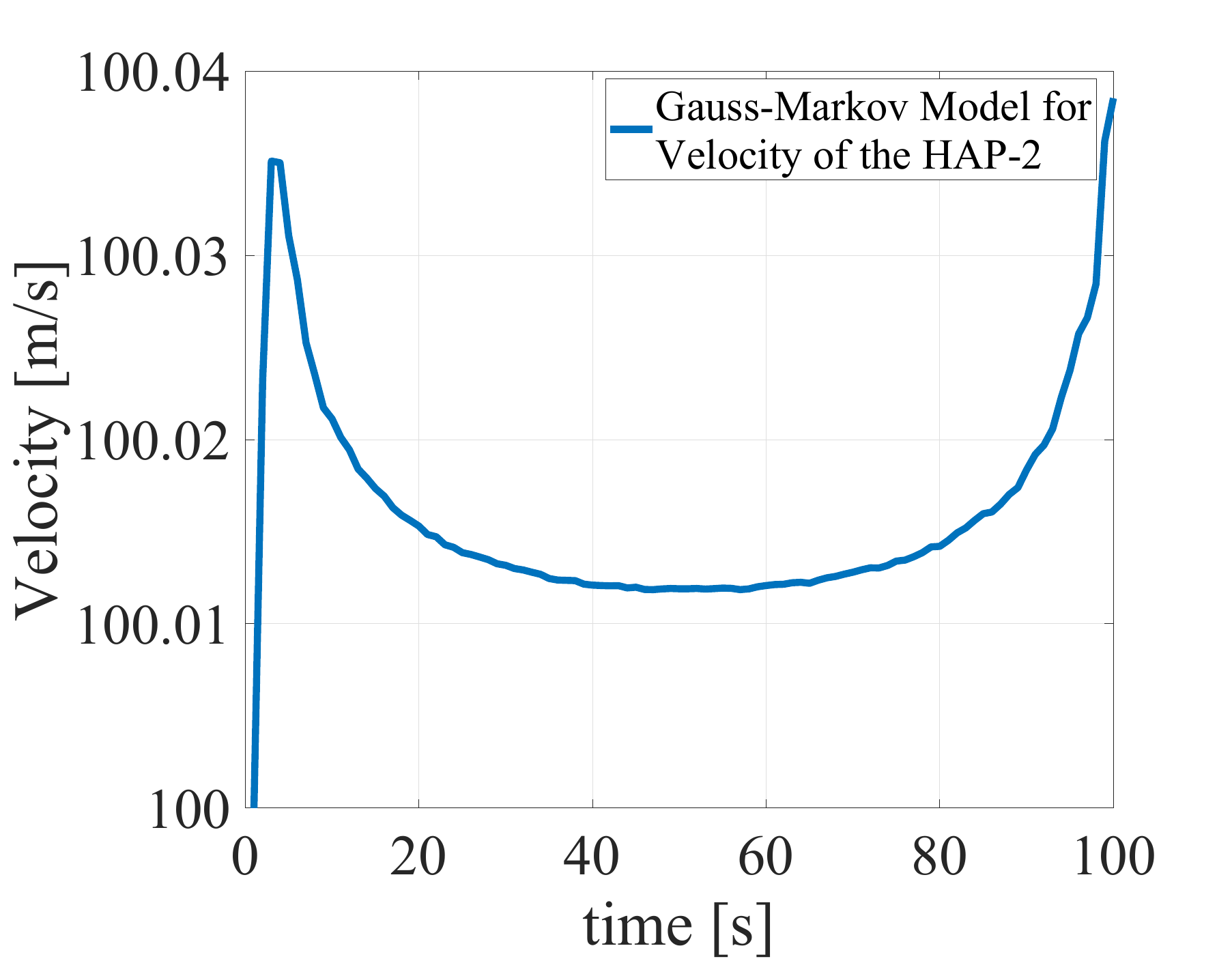}}}%
    \qquad
    \subfigure[Direction vector in azimuth domain]{{\includegraphics[scale=0.18]{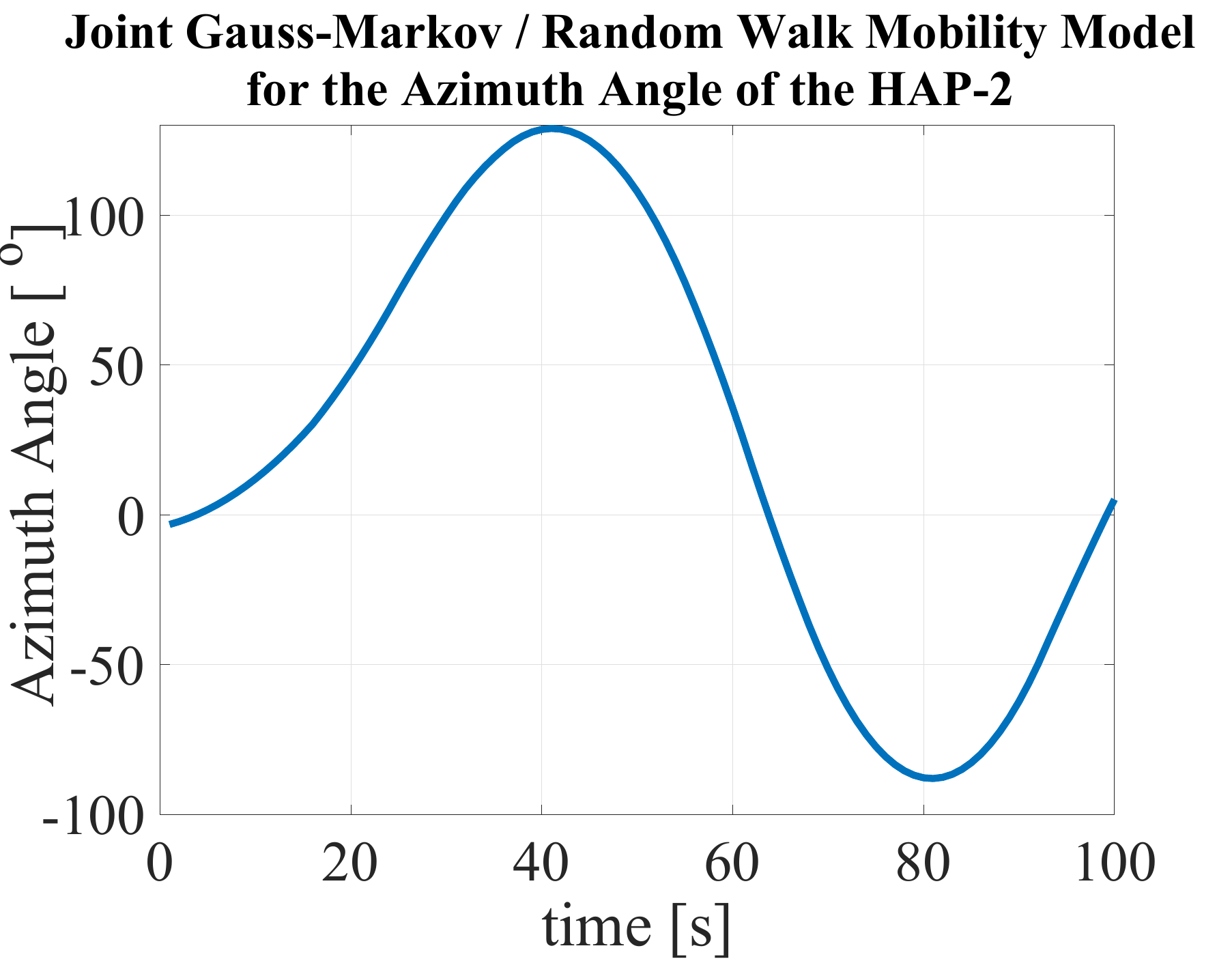}}}%
    \qquad
    \subfigure[Direction vector in elevation domain]{{\includegraphics[scale=0.18]{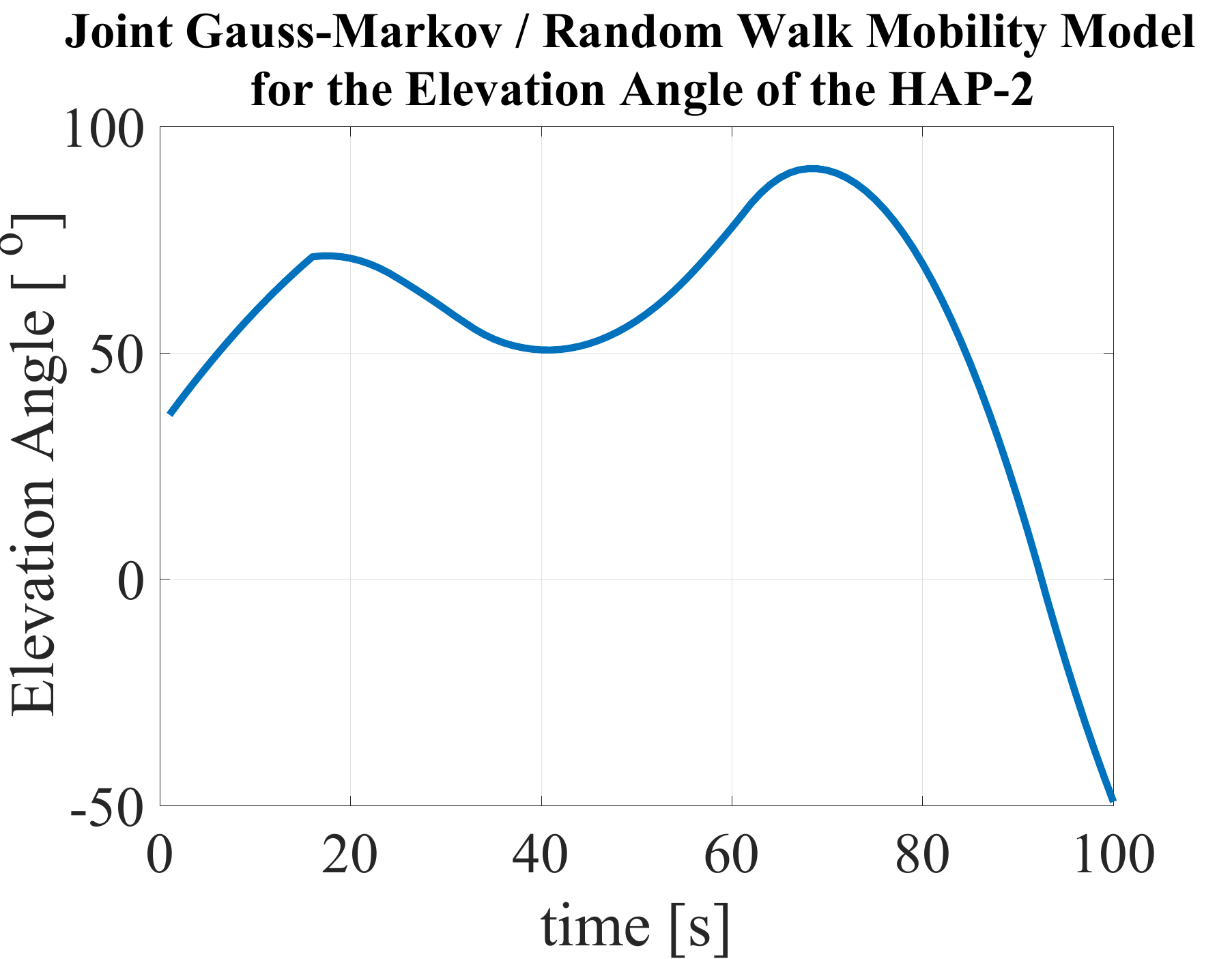}}}%
    \label{vectorsHAP2c}
    \qquad
    \caption{Position vectors of \ac{HAP}-2}%
    \label{vectorsHAP2}%
\end{figure*}

Fig. \ref{vectorsHAP1} and Fig. \ref{vectorsHAP2} show the velocity and direction vectors of \ac{HAP}-1 and \ac{HAP}-2 respectively for a $100$ seconds time duration when the speed of the HAPs is considered as $100$ m/s \cite{ aguiar2008solid}. As it can be seen from Fig. \ref{vectorsHAP1}(a) and Fig. \ref{vectorsHAP2}(a), the variables in the velocity vectors of the two \acp{HAP} are almost equal 100 m/s and does not change drastically during the given time period. The reason why is that for the velocity calculations of the \acp{HAP}, only the Gauss-Markov model is considered with the tuning parameters $\alpha_{\mathcal{V}_{(1,2)}}=\{0.5919, 0.3718\}$. By adopting such a mobility model, it is aimed to keep the velocities of the \acp{HAP} constant rather than changing the values of them in each time iteration.\par

On the other hand, while Fig. \ref{vectorsHAP1}(b) and Fig. \ref{vectorsHAP2}(b) show the direction vectors of \ac{HAP}-1 and \ac{HAP}-2 respectively in the azimuth domain, the direction vectors of \ac{HAP}-1 and \ac{HAP}-2 in the elevation domain are shown in Fig. \ref{vectorsHAP1}(c) and Fig. \ref{vectorsHAP2}(c) respectively. It can be observed from these figures that for both azimuth and elevation directions of both \acp{HAP}, the Gauss-Markov model is preserved until a certain time duration. However, at a randomly selected time, the direction variables are assigned to an irrelevant value that is independent from the previous time instant. Such a mobility model is applied in this study to illustrate the sudden rotations of the \acp{HAP} at an unexpected time instant. \par

\begin{figure}[h]
\centering
\hspace{-0.5cm}
\includegraphics[scale=0.25]{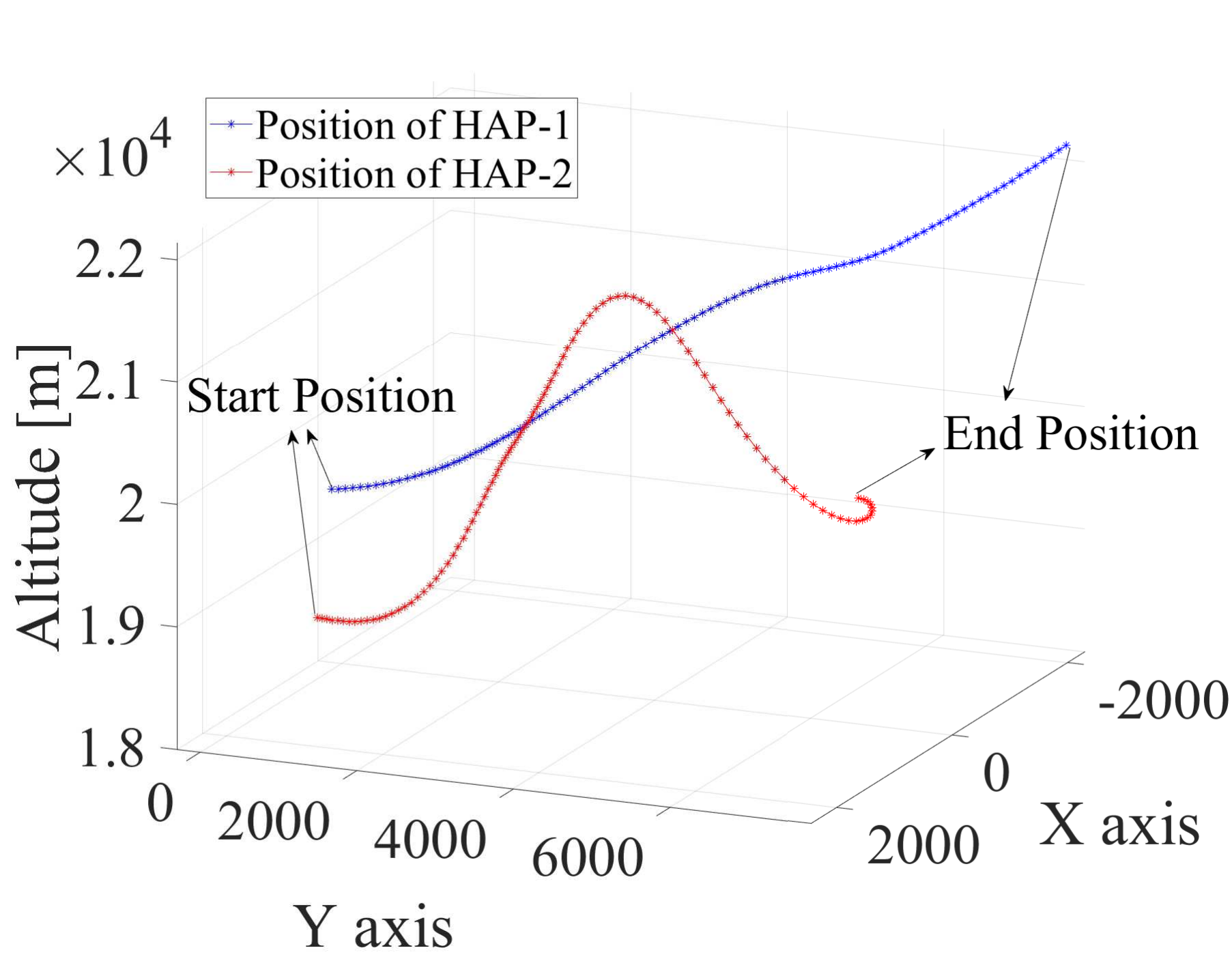}
\caption{Positions of \ac{HAP}-1 and \ac{HAP}-2 in the air for the given time period}
\label{Positions}
\end{figure}

It should be noted that even though the \acp{HAP} that are deployed in real-life scenarios are bulky vehicles and their motion in time is mostly predictable, since their nature allows them to operate in the air by following a scenario based trajectory, such a mixture of Gauss-Markov and 3D random walk model reflects the true mobility characteristics of the \acp{HAP} better than the Gauss-Markov model alone. In order to illustrate the movement of both \acp{HAP} in the air with the joint Gauss-Markov and 3D random walk model, their position change at an altitude $\mathbbm{h}_1\approx\mathbbm{h}_2\approx20$ km in time is given in Fig. \ref{Positions}. In this figure, while the blue points represent the position of \ac{HAP}-1, the red points represent the position of \ac{HAP}-2 in each time iteration. As it can be observed from this figure, while both \acp{HAP} move in a constant direction for a certain time period, at a random moment, directions of them randomly change, which could be justified by a mission update in a real-life scenario. 

\subsection{Effect of Doppler Caused by Mobility}
As stated earlier, in this paper, the communication between two mobile \acp{HAP} is considered to be conducted by a signal that operates at \ac{mmWave} band. Under these conditions, it is expected that the effect of Doppler degrades the quality of communication substantially due to the frequency shift. Therefore, it should be analytically calculated and considered throughout the rest of the proposed channel model. In this regard, the Doppler frequency in terms of speed, direction, and phase angle is given as \cite{ma20193d}

\begin{equation}
\begin{aligned}
\label{DopplerEquation}
    f_d(t_i) =&\frac{\mathcal{V}_1(t_i)}{\lambda}\Biggl(\cos\Bigl(\theta_1\left(t_i\right)-D_{A_1}\left(t_i\right)\Bigl)\\&
    \cos(\phi_1)\cos\Bigl(D_{E_1}\left(t_i\right)\Bigl)+
    \sin(\phi_1)\sin\Bigl(D_{E_1}(t_i)\Bigl)\Biggl)+\\&
    \frac{\mathcal{V}_2(t_i)}{\lambda}\Biggl(\cos\Bigl(\theta_2(t_i)-D_{A_2}(t_i)\Bigl)\\&
    \cos(\phi_2)\cos\Bigl(D_{E_2}(t_i)\Bigl)+
    \sin(\phi_2)\sin\Bigl(D_{E_2}(t_i)\Bigl)\Biggl).
\end{aligned}
\end{equation}

Note that the validity of given equation in (\ref{DopplerEquation}) exists in those cases where the \ac{LoS} condition is satisfied, which means the phase angles of the \acp{HAP} are approximated to be $\theta_1\approx0$, $\theta_2\approx\pi$, and $\phi_1\approx\phi_2\approx\mathcal{D}^{1|2}_{E}$ where $\mathcal{D}^{1|2}_{E}$ is the elevation angle of \ac{HAP}-1 relative to \ac{HAP}-2 and given as 

\begin{equation}
\mathcal{D}^{1|2}_{E}= \tan^{-1}\left(\frac{\mathbbm{h}_1-\mathbbm{h}_2}{D_{2D}}\right).
\end{equation}

However, since the communication between the two \acp{HAP} is in the \ac{mmWave} frequency, the probability of the causes of deviation such as reflection, refraction, etc. is substantially high, which may result in the \ac{LoS} to be broken. In such a case, the given equation in (\ref{DopplerEquation}) has to be adapted to \ac{NLoS} conditions by defining the \ac{AAOD},  \ac{EAOD}, \ac{AAOA}, and \ac{EAOA} of the local scatterers. 

\section{System Analysis}
\label{Sec:Analysis}
\begin{figure*}[h]
    \centering
    \subfigure[Transmitter Structure]{{\includegraphics[scale=0.25]{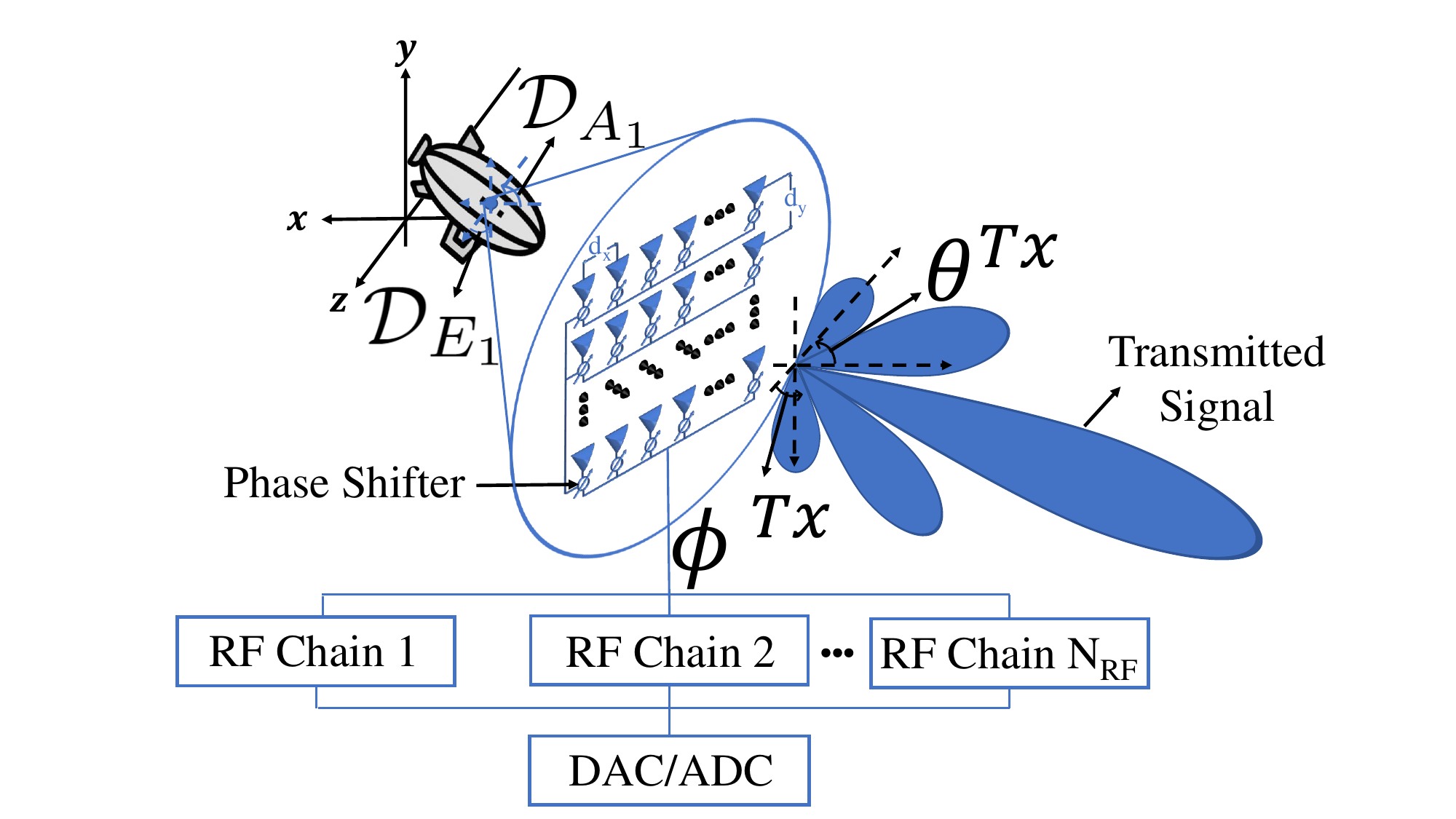}}}%
    \qquad
    \subfigure[Receiver Structure]{{\includegraphics[scale=0.25]{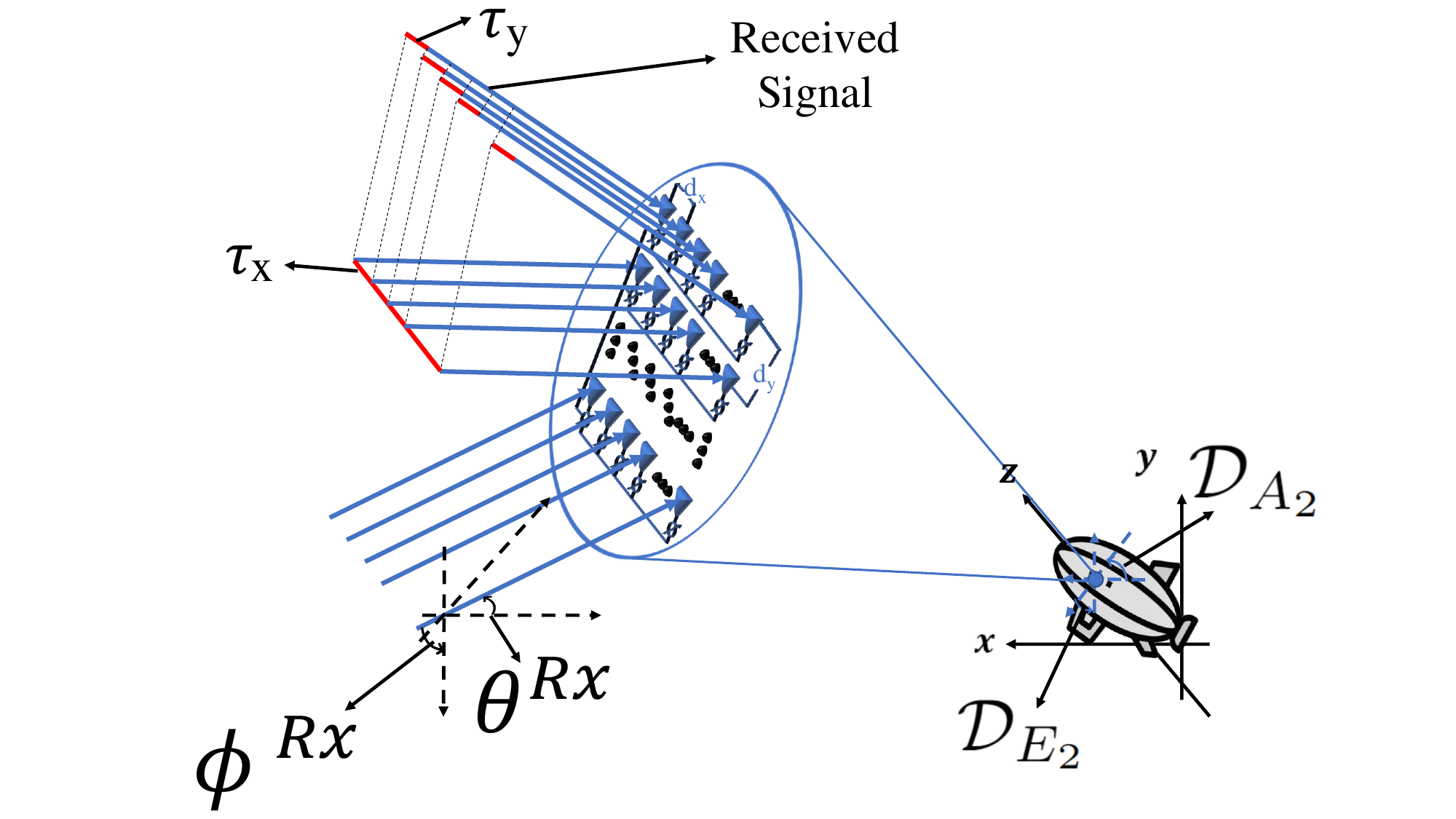}}}%
    \qquad
    \caption{Transmitter and receiver structures of the proposed scenario}%
    \label{UPABeamTx}%
\end{figure*}

In this section, \ac{MIMO} characteristics of the \acp{HAP} are interrogated in terms of beam gain and capacity. In order to operate this interrogation, a signal beam is assumed to be transmitted from the \ac{HAP}-1 and received by the \ac{HAP}-2 at a time instant $t_i$. The hardware characteristics of the \acp{HAP} are considered to be identical where each \ac{HAP} deploys an \ac{URPA} that contains $N\times N$ identical isotropic antennas. Therefore, the number of transmit antennas ($N_{Tx}$) on \ac{HAP}-1 and the number of receive antennas ($N_{Rx}$) on \ac{HAP}-2 are considered to be equal and uniformly placed as shown in Fig. \ref{UPABeamTx}. \par

Under these conditions, the received equivalent signal of a specific antenna element where the signal impinges on the antenna with an angle of arrival in azimuth domain (\ac{AAOA}) and elevation domain (\ac{EAOA}) can be given as 
\begin{comment}
the mathematical formulation of the received signal at time instant $t_i$ can be expressed as 
\begin{equation}
\label{received_vector}
    \mathbf{y(t_i)} = \sqrt{\rho}\mathbf{H(t_i)}\mathbf{x(t_i)}+ \mathbf{\omega(t_i)}
\end{equation}

where $\rho$ is given as the transmit power, $\mathbf{y(t_i)} \in \mathbb{C}^{N_{Rx}\times 1}$ is the received signal, $\mathbf{H(t_i)} \in \mathbb{C}^{N_{Rx}\times N_{Tx}}$ is the channel matrix, $\mathbf{x(t_i)} \in \mathbb{C}^{N_{Tx}\times 1}$ is the transmitted signal, and $\mathbf{\omega(t_i)} \in \mathbb{C}^{N_{Rx}\times 1}$ is the additive white Gaussian noise with zero mean and unity variance. In addition to this expression, the received equivalent signal of a specific antenna element where the signal impinges on the antenna with an angle of arrival in azimuth domain (AAOA) and elevation domain (EAOA) can be given as 
\end{comment}
\begin{equation}
\begin{aligned}
\label{received_mn1}
    y_{m,n}(t_i) & = \sqrt{\rho_{m,n}}h_{m,n}(t_i)\times  s\Bigl(t_i-\big((n-1)\tau_{\text{x}}+ (m-1)\tau_{\text{y}}\big)\Bigl)\\&
    +\omega_{m,n}(t_i),
\end{aligned}
\end{equation}
where $(m,n)$ are the coordination indicators of the antenna array in x-axis and y-axis respectively, and
\begin{equation}
    h_{m,n}(t_i) \in \mathbf{H}(t_i)=\left[\begin{array}{cccc}
h_{1, 1}(t_i) & h_{1, 2}(t_i) & \cdots & h_{1, n}(t_i) \\
h_{2, 1}(t_i) & h_{2, 2}(t_i) & \cdots & h_{2, n}(t_i) \\
\vdots & \vdots & \ddots & \vdots \\
h_{m, 1}(t_i) & h_{m, 2}(t_i) & \cdots & h_{m, n}(t_i)
\end{array}\right]
\end{equation}
is the complex Rayleigh channel coefficient of the $(m,n)$th antenna, $s(\cdot)$ is the transmitted symbol, and
\begin{equation}
    \tau_{\text{x}}=\frac{d_\text{x}\sin{\theta^{Rx}}\cos{\phi^{Rx}}}{c},
\end{equation}
\begin{equation}
  \tau_{\text{y}}=\frac{d_\text{y}\sin{\theta^{Rx}}\sin{\phi^{Rx}}}{c}  
\end{equation}
 are the time delays between the adjacent antennas in x-axis and y-axis respectively with antenna spacing distance $d_{\text{x},\text{y}}=\frac{\lambda_c}{2}$, and $\omega_{m,n}(t_i)$ is the additive white Gaussian noise with zero mean and unity variance. Moreover, by exploiting the time delays in x-axis and y-axis together, the maximum time delay between antenna elements at the receiver can be calculated as 
\begin{equation}
\hspace{-0.25cm}
    \tau_{max}=\left|(N_{Rx}-1)\Biggl(\frac{d_\text{x}\sin{\theta^{Rx}}\cos{\phi^{Rx}}+d_\text{y}\sin{\theta^{Rx}}\sin{\phi^{Rx}}}{c}\Biggl)\right|.
\end{equation}
\par
The representation of the received data can also be given in discrete time by assuming that the amount of Nyquist sampling statistics is adequate. In this case, the regarding representation can be given as 
\begin{equation}
\begin{aligned}
\label{received_mn}
    y_{m,n}(k) =& \sqrt{\rho_{m,n}}\sum_{l=0}^{v-1}h_{m,n}(k,l)\times \\&
    s\Bigl(k-\big((n-1)l+ (m-1)l\big)\Bigl)+\omega_{m,n}(k),
\end{aligned}
\end{equation}
where $h_{m,n}(k,l)$ is given as the channel impulse response of the $l$-th tap and $v$ is the limit of iteration to increase the approximation of finite impulse response of the received data.\par 
If the \ac{DFT} is conducted on (\ref{received_mn}), the received signal of $(m,n)$-th antenna in frequency domain can be calculated as 
\begin{equation}
    y_{m,n}\big(f(t_i)\big)=\mathbf{a}\big({\theta^{Rx},\phi^{Rx},f(t_i)}\big)s\big(f(t_i)\big)+\omega_{m,n}\big(f(t_i)\big),
\end{equation}
where $\mathbf{a}(.)$ is the steering vector and given as \cite{tan2017analysis}
\begin{equation}
    \mathbf{a}\big({\theta,\phi,f(t_i)}\big) = \mathbf{v_x}\big({\theta,\phi,f(t_i)}\big)\otimes\mathbf{v_y}\big({\theta,\phi,f(t_i)}\big).
\end{equation}

The corresponding expressions for $\mathbf{v_x}(.)$ and $\mathbf{v_y}(.)$ are given respectively as 
\begin{equation}
\begin{aligned}
&\mathbf{v_x}\big(\theta,\phi,f(t_i)\big) = \Biggl[1,\hspace{0.20cm}
\exp\left({-j2\pi\frac{f(t_i)}{c}d_\text{x} \sin{\theta} \cos{\phi}}\right),\hspace{0.20cm} \cdots \\& 
\exp\left({-j\left(n-1\right)2\pi\frac{f(t_i)}{c}d_\text{x} \sin{\theta}\cos{\phi}}\right)\Biggl]^T  
 \in \mathbb{C}^
 {1\times N_{Tx,Rx-\text{x}}},
 \end{aligned}
\end{equation}

\begin{equation}
\begin{aligned}
&\mathbf{v_y}\big(\theta,\phi,f(t_i)\big) = \Biggl[1, \hspace{0.20cm}\exp\left({-j2\pi\frac{f(t_i)}{c}d_\text{y} \sin{\theta} \sin{\phi}}\right),\hspace{0.20cm} \cdots \\& 
 \exp\left({-j\left(m-1\right)2\pi\frac{f(t_i)}{c}d_\text{y} ,\sin{\theta}\sin{\phi}}\right)\Biggl]^T \in \mathbb{C}^{1\times N_{Tx,Rx-\text{y}}},
 \end{aligned}
\end{equation}
where  $N_{Tx,Rx-\text{x}}$ are the number of transmit/receive antenna elements in x-axis and $N_{Tx,Rx-\text{y}}$ are the number of transmit/receive antenna elements in y-axis respectively. It should be noted that the steering vector $\mathbf{a}(.)$ is needed to be calculated for both the transmitter \ac{HAP} and receiver \ac{HAP} seperately. In this regard, notations for these two \acp{HAP} can be given as $ \mathbf{a_{Tx}}\big({\theta^{Tx},\phi^{Tx},f(t_i)}\big)$ and $ \mathbf{a_{Rx}}\big({\theta^{Rx},\phi^{Rx},f(t_i)}\big)$ respectively. \par 

Due to the Doppler spread and multipath, the channel between the two \acp{HAP} is expected to be highly selective in both time and frequency domain, which means the channel matrix $\mathbf{H}\big(f(t_i)\big)$ will vary rapidly as the \acp{HAP} move in time. In order to characterize the channel, the steering vectors of the transmitter \ac{HAP} and receiver \ac{HAP} can be exploited. On this matter, representation of the $\mathbf{H}\big(f(t_i)\big)$ for $K$ scattering clusters and $L$ propagation paths can be given as \cite{tse2005fundamentals,el2014spatially} 

\begin{equation}
\begin{aligned}
\label{Hf}
&\mathbf{H}\big(f(t_i)\big)=\frac{N_{Tx}N_{Rx}}{\sqrt{KL}}\sum_{k=1}^K\sum_{l=1}^Lg_{kl}(t_i)\\&
\Lambda_{Rx}\big(\theta_{kl}^{Rx},\phi_{kl}^{Rx},f(t_i)\big)\Lambda_{Tx}\big(\theta_{kl}^{Tx},\phi_{kl}^{Tx},f(t_i)\big)\\&
\mathbf{a_{Rx}}\big(\theta_{kl}^{Rx},\phi_{kl}^{Rx},f(t_i)\big)\mathbf{a^{\text{\textit{H}}}_{Tx}}\big(\theta_{kl}^{Tx},\phi_{kl}^{Tx},f(t_i)\big)\\&
\exp{\left(j2\pi f_{d_{\text{max}}}t_i\sin\theta^{rd}_{kl}\cos\phi^{rd}_{kl}\right)},
 \end{aligned}
\end{equation}
where $g_{kl}$ is the randomly distributed complex Gaussian small-scale  gain of the corresponding $l$-th path of the $k$-th cluster with zero mean and unity variance, $\theta^{Tx}_{kl}$, $\phi^{Tx}_{kl}$, and $\theta^{Rx}_{kl}$  $\phi^{Rx}_{kl}$ are the \ac{AAOD}, \ac{EAOD}, and \ac{AAOA}, \ac{EAOA} of the $l$-th path of the $k$-th cluster respectively, $\Lambda_{Tx}\big(\theta_{kl}^{Tx},\phi_{kl}^{Tx},f(t_i)\big)$ and $\Lambda_{Rx}\big(\theta_{kl}^{Rx},\phi_{kl}^{Rx},f(t_i)\big)$ are the antenna gain for each element of the transmitter and receiver respectively, which are assumed to be $\Lambda_{Tx}\big(\theta_{kl}^{Tx},\phi_{kl}^{Tx},f(t_i)\big)=\Lambda_{Rx}\big(\theta_{kl}^{Rx},\phi_{kl}^{Rx},f(t_i)\big)=1$. \par 
The exponential part in (\ref{Hf}) is usually considered to be $1$ in those cases where the mobility of \acp{HAP} is not considered. However, since in this scenario the mobility plays a crucial role to characterize the channel, the Doppler frequency creates a necessity to consider its maximum value, $f_{d_{\text{max}}}$, in the representation of the channel. It also should be mentioned that the angles $\theta^{rd}_{kl}$ and $\phi^{rd}_{kl}$ used in the exponential part to represent the \ac{AAOA} and \ac{EAOA} are relative to the direction of motion of the receiver \ac{HAP}, which were given in Section \ref{Mobility Model} as $\mathcal{D}_{A_{2}}$ and $\mathcal{D}_{E_{2}}$ for azimuth and elevation domains respectively.

\subsection{Beam Gain}
Due to the fact that the transmitted data is in shape of a narrow beam, the alignment of the signal in accordance with the exact position of the receiver is crucial. In those cases where the beam alignment is not operated successfully, the performance of the communication might either degrade dramatically or get completely lost depending on the severeness of the misalignment of the beam. In \ac{MIMO} systems, the beam alignment is achieved by deploying phase shifters for each analog antenna element to focus the signal on the receiver with a specific angle as shown in Fig. \ref{UPABeamTx}. Even though such an alignment can be applicable on stationary access points, the unpredictable movement of the \acp{HAP} in the proposed scenario creates a challenge to adapt the phase shifters to a different focus angle momentarily. Therefore in this section, the effect of beam misalignment is interrogated. \par

In this regard, the focus angles of the beam at the transmission side is defined as $\theta^{Rx}_F$ and $\phi^{Rx}_F$ in the azimuth and elevation domains respectively. The analog beamforming matrix for UPA is denoted  as 

\begin{equation}
\begin{aligned}
\mathbf{W}=\Biggl[\mathbf{w_1}, \mathbf{w_2}, \cdots, \mathbf{w_{N_{RF}}}\Biggl] \in \mathbb{C}^{N_{Tx}N_{Rx}\times N_{RF}},
\end{aligned}
\end{equation}
where $N_{RF}$ is the number of RF chains at the transmitter and $\mathbf{w}\in \mathbb{C}^{N_{Tx}\times N_{Rx}}$ is the beamforming vector of the corresponding RF chain. \par 
The calculation of the vector $\mathbf{w}$ for each RF chain can be calculated as follows
\begin{equation}
\mathbf{w}=\mathbf{w_x}\otimes \mathbf{w_y}.
\end{equation}
The corresponding expressions for $\mathbf{w_x}(.)$ and $\mathbf{w_y}(.)$ are given respectively as

\begin{equation}
\mathbf{w_x}=\left[e^{-j\beta_{1_\text{x}}}, e^{-j\beta_{2_\text{x}}}, \cdots, e^{-j\beta_{n_\text{x}}}\right] \in \mathbb{C}^{1\times N_{Tx-\text{x}}},
\end{equation}
\begin{equation}
\mathbf{w_y}=\left[e^{-j\beta_{1_\text{y}}}, e^{-j\beta_{2_\text{y}}}, \cdots, e^{-j\beta_{m_\text{y}}}\right]  \in \mathbb{C}^{1\times N_{Tx-\text{y}}},
\end{equation}
where 
\begin{equation}
    \beta_{n_\text{x}}= \frac{2\pi}{\lambda_c}(n-1)d_\text{x}\sin{\theta^{Tx}}\cos{\phi^{Tx}},
\end{equation}
\begin{equation}
    \beta_{m_\text{y}}= \frac{2\pi}{\lambda_c}(m-1)d_\text{y}\sin{\theta^{Tx}}\sin{\phi^{Tx}}
\end{equation}
are the phase shifters of the ($m,n$)-th antenna element.\par

By utilizing the beamforming matrix $\mathbf{W}$ and the steering vector of the receiver $\mathbf{a_{Rx}}\big({\theta^{Rx},\phi^{Rx},f(t_i)}\big)$, the beam gain can be calculated as follows\footnote{Note that the transmitter gain can also be obtained by exploiting (\ref{gainraw}) by changing the indicator "$\textit{Rx}$" to "$\textit{Tx}$"}
\begin{equation}
\begin{aligned}
\label{gainraw}
&g\big(\mathbf{W},\theta^{Rx},\phi^{Rx},f(t_i)\big)= \mathbf{W}^{\text{\textit{H}}}\mathbf{a_{Rx}} \big({\theta^{Rx},\phi^{Rx},f(t_i)}\big) \\ &=\sum_{n=1}^{N_{Tx-\text{x}}}\sum_{m=1}^{N_{Tx-\text{y}}}e^{-j2\pi(n-1)\frac{d_\text{x}}{c}\mu_\text{x}}\otimes e^{-j2\pi(m-1)\frac{d_\text{y}}{c}\mu_\text{y}}, 
 \end{aligned}
\end{equation}
which can be further expressed as
\begin{equation}
\begin{aligned}
\label{beamgain}
   g(\mu_{\text{x}},\mu_{\text{y}},t_i) =& \frac{\sin\left(-N_{Tx}\pi\frac{d_{\text{x}}}{c}\mu_{\text{x}}\right)}{\sin\left(-\pi\frac{d_{\text{x}}}{c}\mu_{\text{x}}\right)}e^{-2j\pi\frac{d_{\text{x}}}{c}(N_{Tx}-1)\mu_{\text{x}}}\otimes \\& \frac{\sin\left(-N_{Tx}\pi\frac{d_{\text{y}}}{c}\mu_{\text{y}}\right)}{\sin\left(-\pi\frac{d_{\text{y}}}{c}\mu_{\text{y}}\right)}e^{-2j\pi\frac{d_{\text{y}}}{c}(N_{Tx}-1)\mu_{\text{y}}},
\end{aligned}
\end{equation}
where
\begin{equation}
    \mu_{\text{x}}=\big(f_c(t_i)+f_d(t_i)\big)\sin\theta^{Rx}-f_c(t_i)\sin\theta^{Rx}_F,
\end{equation}
\begin{equation}
    \mu_{\text{y}}=\big(f_c(t_i)+f_d(t_i)\big)\sin\phi^{Rx}-f_c(t_i)\sin\phi^{Rx}_F.
\end{equation}

The maximum achievable gain can be obtained as $g_{\text{max}}=g(0,0,t_i)=N_{Tx}$ in the case where $\mu_{\text{x}}=\mu_{\text{y}}=0$, which means the effect of Doppler does not exist and the focus angles in both azimuth and elevation domains that are required to impinge the transmitted signal upon the receiver are equal to the arrival angles of the beam. In this case, the effective channel between the transmitter and receiver can be interpreted to be the multiplication of the analog beamforming matrix and the steering vector of the receiver, which can be mathematically expressed as                          
\begin{equation}
    h_{eff}= g\big(\mathbf{W},\theta^{Rx},\phi^{Rx},f(t_i)\big)=\mathbf{W}^{\text{\textit{H}}}\mathbf{a_{Rx}}\big({\theta^{Rx},\phi^{Rx},f(t_i)}\big).
\end{equation}
However, when the effect of Doppler is included, the matching possibility of the arrival angles and the focus angles decreases dramatically due to the dispersion in angular domain, which causes the focus angles to get shifted from their desired values. In such a case, the phase shifting is required to be operated in such a way that the analog beamforming matrix can compensate the effect of the beam-selective channel.

\begin{figure}[h]
\centering
\includegraphics[scale=0.45]{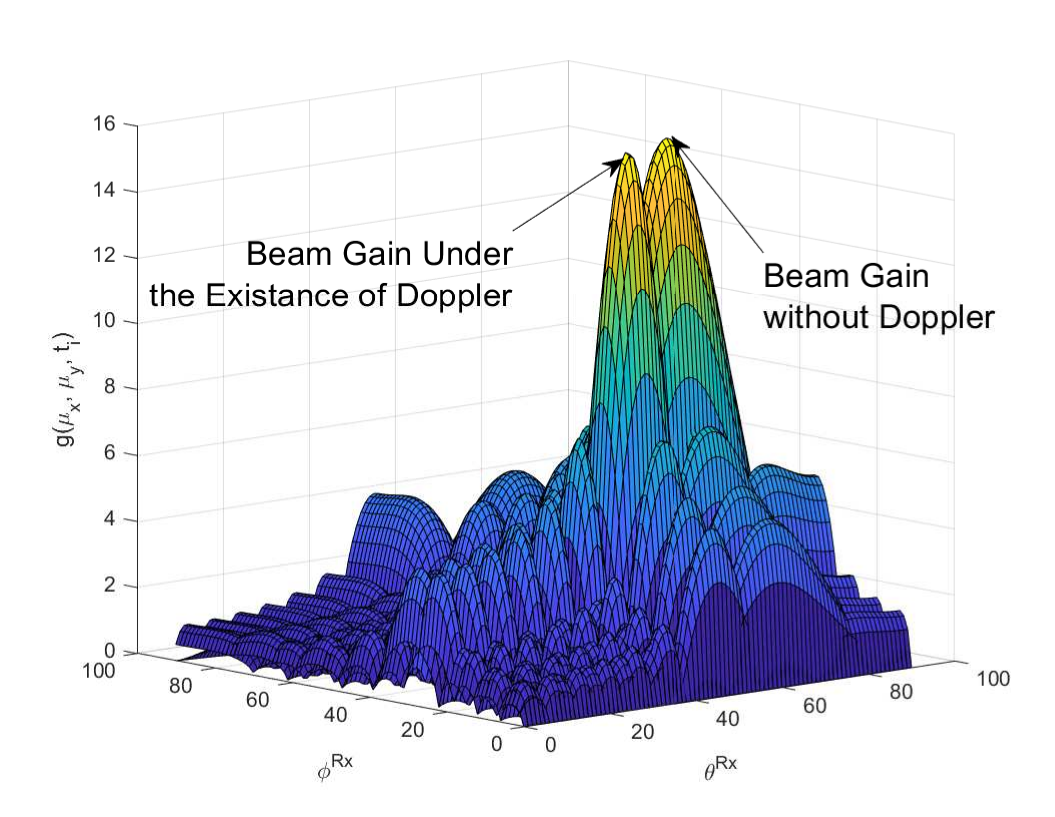}
\caption{Beam gain for $N_{Tx}=16$, and $(\theta^{Rx}_F,\phi^{Rx}_F)=(60^o,30^o)$}
\label{gainNoDoppler}
\end{figure}

The illustration of the beam gain at a time instant $t_i$ is given in Fig. \ref{gainNoDoppler} for $N_{Tx}=16$ and $(\theta^{Rx}_F,\phi^{Rx}_F)=(60^o,30^o)$. As it can be seen from this figure, in the case where the Doppler doesn't exist, the beam gain reaches its maximum value $N_{Tx}$ when the direction angles of the arriving beam $(\theta^{Rx},\phi^{Rx})$ are equal to the selected focus angles $(\theta^{Rx}_F,\phi^{Rx}_F)$ and gradually degrades as the difference between $(\theta^{Rx},\phi^{Rx})$ and $(\theta^{Rx}_F,\phi^{Rx}_F)$ gets larger. However, when the Doppler is included, the focus angles are shifted and the maximum gain is not achieved on the desired values.\par

\subsection{System Capacity}
Due to the high mobility of the \acp{HAP} in the proposed scenario, it is expected that the Doppler spread causes the channel to be highly selective in time domain and multipath causes the channel to be highly selective in frequency domain. In order to compensate the effects of the frequency-selective channel, it's assumed that a \ac{CP} that is longer than the time delay is utilized in time domain to avoid \ac{ISI}. Therefore, the calculations of the system capacity is conducted by considering only the existence of time-selective channel. Under the existence of such a time-selective channel, when a multicarrier waveform is utilized rather than a single-carrier waveform, the resulting interference between the carriers has the potential to decrease the performance of the communication due to the fact that the orthogonality between these carriers is corrupted. In order to evaluate the effect of the interference between these carriers, namely \ac{ICI}, upon the multicarrier communication between the two mobile \acp{HAP}, this section analyses the \ac{SINR} performance of the proposed scenario.\par

In this manner, the received signal of ($m,n$)-th antenna calculated in (\ref{received_mn}) can be transformed into 
\begin{equation}
\label{received_mnp}
\begin{aligned}
    &y_{m,n}\big(p_{N_c},f(t_i)\big)  =   \sqrt{\rho_{m,n}} \bigg(h_{m,n}\big(p_{N_c},f(t_i)\big)\times \\&
    s\Bigl(p_{N_c},t_i-\big((n-1)\tau_\text{x}+(m-1)\tau_\text{y}\big)\Bigl)+ \\&
    \sum_{\substack{q=0 \\ q\neq p}}^{N_c-1}h_{m,n}(q,f(t_i))\times \\&
    s\Bigl(q,t_i-\big((n-1)\tau_\text{x}+(m-1)\tau_\text{y}\big)\Bigl)\bigg)+\omega_{m,n}(p,t_i),     
\end{aligned}
\end{equation}
where $N_c$ is the number of carriers in each antenna element, $p\in \{0, 1, \cdots, N_c-1\}$ is the specific carrier of the ($m,n$)-th antenna element that the data is received from, and $q\in\{0, 1, \cdots, N_c-1\}, q\neq p$ is all the rest of the carriers except $p$ that are considered as noise due to the \ac{ICI}. In order to take (\ref{received_mnp}) one step further, the case where the spatial diversity is exploited can be considered. In such a case the received signal formula evolves into 

\begin{equation}
\label{received_mnp_vector}
\begin{aligned}
   & \mathbf{y_{m,n}}(f(t_i))  =  \sqrt{\rho_{m,n}}  \bigg (\bigg[ h_{m,n}\big(p_{1},p_{1},f(t_i)\big)\times  s\Bigl(p_1,t_i-\\&\big((n-1)\tau_\text{x}+(m-1)\tau_\text{y}\big)\Bigl), \cdots,  h_{m,n}\big(p_{N_c},p_{N_c},f(t_i)\big)\times\\&  s\Bigl(p_{N_c},t_i-\big((n-1)\tau_\text{x}+(m-1)\tau_\text{y}\big)\Bigl) \bigg]+\\& \hspace{-5pt}\sum_{\substack{q=0\\ q\neq p}}^{N_c-1}h_{m,n}\big(p,q,f(t_i)\big)\times  s\Bigl(q,t_i-\big((n-1)\tau_\text{x}+(m-1)\tau_\text{y}\big)\Bigl) \bigg)\\&+
   \mathbf{\omega_{m,n}}(p_{N_c},t_i) \in \mathbb{C}^{1\times N_{c}},     
\end{aligned}
\end{equation}
where 
\begin{equation}
\begin{aligned}
    &\mathbf{h_{m,n}}\big(p_{N_c},f(t_i)\big)\in [h_{m,n}\big(p_1,f(t_i)\big), \cdots, h_{m,n}\big(p_{N_c},f(t_i)\big)],\\& \mathbb{C}^{1\times N_c}
 \end{aligned}
\end{equation}
is the channel vector of the ($m,n$)-th antenna element that consists of the channel coefficients of each carrier $h_{m,n}\big(p_{N_c},f(t_i)\big)$.
For the sake of simplification, (\ref{received_mnp_vector}) can be reformed as 

\begin{equation}
\hspace{-10pt}
\begin{aligned}
   & \mathbf{y_{m,n}}(f(t_i)) = \sqrt{\rho_{m,n}}\bigg(\mathbf{h_{m,n}}\big(p_{N_c},p_{N_c},f(t_i)\big) \times \\&
    \mathbf{x}\Bigl(p_{N_c},t_i-\big((n-1)\tau_\text{x}+(m-1)\tau_\text{y}\big)\Bigl) + \sum_{\substack{q=0 \\ q\neq p}}^{N_c-1} \mathbf{h_{m,n}}\big(p,q,f(t_i)\big) \\&
    \mathbf{x}\Bigl(q,t_i-\big((n-1)\tau_\text{x}+(m-1)\tau_\text{y}\big)\Bigl)\bigg) + \mathbf{\omega_{m,n}}(p_{N_c},t_i) \in \mathbb{C}^{1\times N_{c}}.
 \end{aligned}
\end{equation}
In this case, the received signal at time instant $t_i$ can be expressed in vector form as
\begin{equation}
\begin{aligned}
\mathbf{Y}(f(t_i)) =[&\mathbf{y_{1, 1}}(f(t_i)), \mathbf{y_{1, 2}}(f(t_i)), \cdots, \mathbf{y_{1, n}}(f(t_i)), \\& 
        \mathbf{y_{2, 1}}(f(t_i)), \mathbf{y_{2, 2}}(f(t_i)), \cdots, \mathbf{y_{2, n}}(f(t_i)), \cdots, \\ &
        \mathbf{y_{m, 1}}(f(t_i)), \mathbf{y_{m, 2}}(f(t_i)), \cdots, \mathbf{y_{m, n}}(f(t_i))], \\ &
     \in \mathbb{C}^{mn\times N_{c}},
 \end{aligned}
\end{equation}
and in closed form as 
\begin{equation}
\label{received_closed_form}
    \mathbf{Y}(f(t_i))=\mathbf{H}\big(p, p, f(t_i)\big) \mathbf{X}(p)+\sum_{\substack{q=0 \\ q \neq p}}^{N_c-1} \mathbf{H}\big(p, q, f(t_i)\big) \mathbf{X}(q)+\mathbf{\omega}(p).
\end{equation}
The \ac{ICI} analysis of the proposed multicarrier \ac{HAP} communication can be based upon (\ref{received_closed_form}) by calculating the covariance matrix of the total noise $\mathbf{\hat{\omega}}(p) = \mathbf{\omega_{\mathrm{ICI}}}(p)+\mathbf{\omega}(p)$ where 
\begin{equation}
 \mathbf{\omega_{\mathrm{ICI}}}(q)=\sum_{\substack{q=0 \\ q \neq p}}^{N_c-1} \mathbf{H}\big(p, q, f(t_i)\big) \mathbf{X}(q).
 \end{equation}
 The procedure to mathematically express the covariance of  $\mathbf{\omega_{ICI}}(p)$ can be represented as 
 
 \begin{equation}
 \label{covariance_ICI}
     \begin{aligned}
        & \mathbb{E}\left[\mathbf{\omega_{\mathrm{ICI}}}(q) \mathbf{\omega_{\mathrm{ICI}}}^H(q)\right]=\\& 
        \mathbb{E}\left[\sum_{\substack{p=0 \\ q \neq j}}^{N_c-1} \sum_{\substack{q=0 \\ q \neq j}}^{N_c-1} \mathbf{H}\big(j, p ,f(t_i)\big) \mathbf{X}(p) \mathbf{X}^H(q) \mathbf{H}^H\big(j, q ,f(t_i)\big)\right] \\&
         =\sum_{\substack{p=0 \\ q \neq j}}^{N_c-1} \sum_{\substack{q=0 \\ q \neq j}}^{N_c-1} \mathbb{E}\left[\mathbf{H}\big(j, p ,f(t_i)\big) \mathbf{R}_{X X}(p, q) \mathbf{H}^H\big(j, q ,f(t_i)\big)\right],
     \end{aligned}
 \end{equation}
 where $\mathbf{R}_{X X}(p, q)$ is defined to represent the term $\mathbb{E}\left[\mathbf{X}(p)\mathbf{X}(q)\right]$. \par
 The following analysis are composed on the term $\mathbb{E}\left[\mathbf{H}\big(j, p ,f(t_i)\big) \mathbf{R}_{X X}(p, q) \mathbf{H}^H\big(j, q ,f(t_i)\big)\right]$ given in (\ref{covariance_ICI}) for the sake of simplicity. In this regard, we start our analysis with the  eigendecomposition of the signal covariance matrix given as \cite{umeyama1988eigendecomposition} 

 \begin{equation}
 \label{eigendecomposition}
     \mathbf{R}_{X X}(p, q)= \mathbf{U}(p, q) \boldsymbol{\Lambda}_X(p, q) \mathbf{U}^H(p, q),
 \end{equation}
 where
 $\mathbf{U}(p, q)$ is the unitary matrix that contain the eigenvectors and
 \begin{equation}
 \label{diag}
 \boldsymbol{\Lambda}_X(p, q)=\text{diag}\big(\lambda_1(p,q), \cdots, \lambda_{N_{Tx}}(p,q)\big)  
 \end{equation}
 is the diagonal matrix of the eigenvalues of $\mathbf{R}_{X X}(p, q)$. In the conducted analysis, it is assumed that the channel matrix $\mathbf{H}\big(j, p ,f(t_i)\big)$ consists of independent and identically distributed Gaussian variables. Therefore, it can be stated that the distribution of  $\mathbf{H}\big(j, p ,f(t_i)\big)$ is equivalent to $\mathbf{H}\big(j, p ,f(t_i)\big)\mathbf{U}(p, q)$. In this case, the term  $\mathbb{E}\left[\mathbf{H}(j, p) \mathbf{R}_{X X}(p, q) \mathbf{H}^H\big(j, q ,f(t_i)\big)\right]$ can be reformed as 
 \begin{equation}
 \label{expectedWithLambda}
     \begin{aligned}
         &\mathbb{E}\left[\mathbf{H}\big(j, p ,f(t_i)\big) \mathbf{R}_{X X}(p, q) \mathbf{H}^H\big(j, q ,f(t_i)\big)\right]= \\&
         \mathbb{E}\left[\mathbf{H}\big(j, p ,f(t_i)\big) \boldsymbol{\Lambda}_X(p, q) \mathbf{H}^H\big(j, q ,f(t_i)\big)\right].
     \end{aligned}
 \end{equation}
 \par
 The investigation of the matrix given in (\ref{expectedWithLambda}) can be conducted more thoroughly by choosing some specific indices. In this regard, the matrix can be represented for the ($r,s$)-th element as 
 \begin{equation}
 \label{expected_rs}
 \begin{aligned}
   & \{\mathbb{E}\left[\mathbf{H}\big(j, p ,f(t_i)\big) \boldsymbol{\Lambda}_X(p, q) \mathbf{H}^H\big(j, q ,f(t_i)\big)\right]\}_{r,s}=\\& 
     \sum_{k=1}^{N_{Tx}}\lambda_k(p,q)\mathbb{E}\left[\{\mathbf{H}\big(j, p ,f(t_i)\big)\}_{r,k}\{\mathbf{H}^H\big(j, q ,f(t_i)\big)\}_{k,s}\right]=\\& 
     \sum_{k=1}^{N_{Tx}}\lambda_k(p,q)\mathbb{E}\left[\{\mathbf{H}\big(j, p ,f(t_i)\big)\}_{r,k}\{\mathbf{H}^H\big(j, q ,f(t_i)\big)\}_{k,r}\right]\delta_{r-s},
 \end{aligned}
 \end{equation}
 where $\delta_{r-s}$ is the correlation parameter of the matrix. \par
 
 Note that the final form of (\ref{expected_rs}) is only valid for the case where the channel is uncorrelated, which means that when $r\neq s$, the cross-correlation between these values becomes zero and leads the (\ref{expected_rs}) to have a non-zero value. In the case where $r=s$ the auto-correlation becomes maximum and leads the (\ref{expected_rs}) to become zero. \par 

 By using the Fourier basis, (\ref{expected_rs}) can be represented as

 \begin{equation}
     \begin{aligned}
         &\mathbb{E}\left[\{\mathbf{H}\big(j, p ,f(t_i)\big)\}_{r,k}\{\mathbf{H}^H\big(j, q ,f(t_i)\big)\}_{k,r}\right] = \\&
         \frac{1}{{N_c}^2}\left(\sum_{l=0}^{v-1}\exp\big(-j2\pi l(p-q)/N_c\big)\right)\times \\&
         \sum_{r_1=0}^{N_c-1}\sum_{r_2=0}^{N_c-1}  \mathbb{E}\left[h_{m,n}^{(k,r)}(r_1,l)h_{m,n}^{(k,r)}(r_2,l)\right] \times \\&
         \exp\big(j2\pi r_1(p-j)/N_c\big) \exp\big(-j2\pi r_2(q-j)/N_c\big).
     \end{aligned}
 \end{equation}
 
 As a result, the covariance matrix of the \ac{ICI} can be given as  
 \begin{equation}
     \begin{aligned}
        &\mathbb{E}\left[\mathbf{\omega_{\mathrm{ICI}}}(q) \mathbf{\omega_{\mathrm{ICI}}}^H(q)\right]= \\&
        \sum_{\substack{p=0 \\ q \neq j}}^{N_c-1} \sum_{\substack{q=0 \\ q \neq j}}^{N_c-1}\frac{\text{tr}\big(\boldsymbol{\Lambda}_X(p, q)\big)}{{N_c}^2} \left(\sum_{l=0}^{v-1}\exp\big(-j2\pi l(p-q)/N_c\big)\right)
        \times \\&
        \Biggl( \sum_{r_1=0}^{N_c-1}\sum_{r_2=0}^{N_c-1}\mathbb{E}\left[h_{m,n}^{(k,r)}(r_1,l)h_{m,n}^{(k,r)}(r_2,l)\right]\\&
        \exp\big(j2\pi r_1(p-j)/N_c\big) 
        \exp\big(-j2\pi r_2(q-j)/N_c\big) \Biggl).
     \end{aligned}
 \end{equation}
\par

After obtaining the covariance matrix of \ac{ICI}, the calculation of \ac{SINR} at the $p$-th carrier where $1\leq p\leq N_c$ can be operated as follows 
\begin{equation}
\label{SINR}
    \gamma_p=\frac{E_{x,p}\left|\mathbf{H}\big(p, p ,f(t_i)\big)\right|^2}{\sum_{q\neq p}E_{x,q}\left|\mathbf{H}\big(p, q ,f(t_i)\big)\right|^2+E_{x,\omega}\left|\omega(t_i)\right|},
\end{equation}
where $E_{x,p}$ is the input energy that is allocated to the $p$-th carrier and  $E_{x,q}$ is the leakage energy from the $q$th carrier that contaminates the $p$-th carrier.\par
Since the analysis of \ac{SINR} is achieved, now the ergodic capacity of the system can be calculated as follows
\begin{equation}
    C=\sum_{p=1}^{N_c} \log_2\left(1+\gamma_p\right).
\end{equation}
\section{PDF of the Signal to Interference Plus Noise Ratio} 
\label{Sec:pdf}
In this section, the analytical derivation of the \ac{PDF} for the proposed channel model's \ac{SINR} is presented. In this regard, the beam gain given in (\ref{beamgain}) is approximated for more traceability as follows \cite{yu2017coverage}
\begin{equation}
\label{beamgainapprox}
\hspace{-3pt}
  g(\theta^{Rx},\phi^{Rx})=\cos^2\left(\frac{\pi {N_{Tx-\text{x}}}}{2} \theta^{Rx}\right)\otimes \cos^2\left(\frac{\pi {N_{Tx-\text{y}}}}{2} \phi^{Rx}\right). 
\end{equation}
 Note that the given formulation of $g(\theta^{Rx},\phi^{Rx})$ is valid only in the case where the $|\theta^{Rx}| \leq  \frac{1}{{N_{Tx-\text{x}}} }$, and $|\phi^{Rx}| \leq \frac{1}{{N_{Tx-\text{y}}} }$. If these conditions are not satisfied, the beam gain is considered to be $0$. %The reason why such a cosine based beam gain formulation is adopted in the derivation of \ac{PDF} is that the beam gain given in (\ref{beamgainapprox}) has the capability to provide a decent approximation for the gain of the main lobe while sacrificing the accuracy of the beam gain of the side lobes, which is neglected in the \ac{PDF} analysis of the \ac{SINR} for the sake of decreasing the complexity.
 \par

To further simplify the beam gain formulation, (\ref{beamgainapprox}) can be rewritten as follows
\begin{equation}
\begin{aligned}
& g(\theta^{Rx},\phi^{Rx}) \simeq  N_{Tx-\text{x}} \Pi\left(N_{Tx-\text{x}} \theta^{Rx}\right)+\\&
N_{Tx-\text{y}} \Pi\left( N_{Tx-\text{y}} \phi^{Rx}\right)+\\& N_{Tx} \cos^2 \left(\frac{\pi N_{Tx-\text{x}}}{2}\right) \cos^2 \left(\frac{\pi N_{Tx-\text{y}}}{2}\right) \times \\&
\Pi\left({N_{Tx-\text{x}}\left|\theta^{Rx}\right|}\right)\times\Pi\left({N_{Tx-\text{y}}\left|\phi^{Rx}\right|}\right),
\end{aligned}
\end{equation}
where $\Pi(.)=\left\{\begin{array}{lll}1 & \text { for } & |.| \leq 1 \\ 0 & \text { for } & |.|>1\end{array}\right.$.
By referring the fact that the $\{\theta^{Rx},\phi^{Rx}\}$ are Gaussian distributed, the \ac{PDF} of the $g(\theta^{Rx},\phi^{Rx})$ can be given as 
\begin{equation}
\label{PDFgRx}
\begin{aligned}
&f_{g(\theta^{Rx},\phi^{Rx})}\left(g(\theta^{Rx},\phi^{Rx})\right)=A_{{Rx}}\left(\theta^{Rx^{\prime}}, \phi^{Rx}\right) \times\\&
 \delta\left(g(\theta^{Rx},\phi^{Rx})-N_{Tx} \cos^2 \left(\frac{\pi N_{Tx-\text{x}}}{2}\right) 
\cos^2 \left(\frac{\pi N_{Tx-\text{y}}}{2}\right)\right),
\end{aligned}
\end{equation}
where $\delta(.)$ is the Dirac delta function and $A_{{Rx}}$ is the derivation variable given as
\begin{equation}
\begin{aligned}
&A_{{Rx}}\left(\theta^{Rx^{\prime}}, \phi^{Rx}\right) = 
Q\left(\frac{N_{Tx}  \theta^{Rx^{\prime}}}{N_{Tx}  \phi^{Rx}}\right)- \\&
\left(Q\left(\frac{1+N_{Tx}  \theta^{Rx^{\prime}}}{N_{Tx}  \phi^{Rx}}\right)+  
Q\left(\frac{1-N_{Tx}  \theta^{Rx^{\prime}}}{N_{Tx}  \phi^{Rx}}\right)\right),
\end{aligned}
\end{equation}
by exploiting the $Q$-$function$, $Q(.)$.\par

Referring to (\ref{gainraw}), the instantaneous directivity gain of the proposed model can be given as 

\begin{equation}
\label{directivityGain}
    \begin{aligned}
\mathbb{G}\left(\theta^{Tx/Rx},\phi^{Tx/Rx}\right)= &
g\left(\mathbf{W},\theta^{Tx},\phi^{Tx},f(t_i)\right) \\&
g\left(\mathbf{W},\theta^{Rx},\phi^{Rx},f(t_i)\right).
    \end{aligned}
\end{equation}

By exploiting (\ref{PDFgRx}) and (\ref{directivityGain}), the \ac{PDF} of directivity gain conditioned on the receiver array gain can be given as 

\begin{equation}
\label{PDFG}
\begin{aligned}
& f_{\mathbb{G} \mid g(\theta^{Rx},\phi^{Rx})}(\mathbb{G})=\frac{A_{{Tx}}\left(\theta^{Tx^{\prime}}, \phi^{Tx}\right)}{g(\theta^{Rx},\phi^{Rx})} \times \\&
  \delta\left(\frac{\mathbb{G}}{g(\theta^{Rx},\phi^{Rx})}-N_{Tx} \cos^2 \left(\frac{\pi N_{Tx-\text{x}}}{2}\right) 
\cos^2 \left(\frac{\pi N_{Tx-\text{y}}}{2}\right)\right) .
\end{aligned}
\end{equation}

From (\ref{PDFgRx}) and (\ref{PDFG}), derivation of the \ac{PDF} for the directivity gain $\mathbb{G}$ is given as follows

\begin{equation}
\begin{aligned}
&f_{\mathbb{G}}(\mathbb{G})=   \int f_{\mathbb{G} \mid g(\theta^{Rx},\phi^{Rx})}(\mathbb{G})\times \\& f_{g(\theta^{Rx},\phi^{Rx})}\left(g(\theta^{Rx},\phi^{Rx})\right) d g(\theta^{Rx},\phi^{Rx}) = \\&
 \int \frac{A_{{Rx}}\left(\theta^{Rx^{\prime}}, \phi^{Rx}\right) A_{{Tx}}\left(\theta^{Tx^{\prime}}, \phi^{Tx}\right)}{g(\theta^{Rx},\phi^{Rx})}\times  \\
& \delta\left(g(\theta^{Rx},\phi^{Rx})-N_{Tx}\cos^2 \left(\frac{\pi N_{Tx-\text{x}}}{2}\right)\cos^2 \left(\frac{\pi N_{Tx-\text{y}}}{2}\right)\right) \\
& \delta\left(\frac{\mathbb{G}}{g(\theta^{Rx},\phi^{Rx})}- N_{Tx}\cos^2 \left(\frac{\pi N_{Tx-\text{x}}}{2}\right)\cos^2 \left(\frac{\pi N_{Tx-\text{y}}}{2}\right)\right)\\&
dg(\theta^{Rx},\phi^{Rx}),
\end{aligned}
\end{equation}

\begin{equation}
\label{PDFGClosedForm}
\begin{aligned}
f_{\mathbb{G}}(\mathbb{G})= & \frac{A_{{Rx}}\left(\theta^{Rx^{\prime}}, \phi^{Rx}\right) A_{{Tx}}\left(\theta^{Tx^{\prime}}, \phi^{Tx}\right)}{\kappa(N_{Tx-\text{x}}, N_{Tx-\text{y}})}  \times \\&
\delta\left(\mathbb{G}-\kappa(N_{Tx-\text{x}}, N_{Tx-\text{y}})\right),
\end{aligned}
\end{equation}
where 
\begin{equation}
    \begin{aligned}
    &\kappa(N_{Tx-\text{x}}, N_{Tx-\text{y}})=  N_{Tx}^2\cos^4 \left(\frac{\pi N_{Tx-\text{x}}}{2}\right)\cos^4 \left(\frac{\pi N_{Tx-\text{y}}}{2}\right).
    \end{aligned}
\end{equation}

Now, let us define a Nakagami random variable notated as $\mathbbm{r}$ and $\zeta=\mathbbm{r}^2$. Then, the normalized Gamma random variable can be given as 

\begin{equation}
\label{RV}
f_\zeta(\zeta)=\frac{(m \zeta)^{m}}{\Gamma(m)} \exp (-m \zeta), \quad \zeta>0
\end{equation}
where $m$ is the Nakagami fading parameter and $\Gamma(.)$ is the Gamma function \cite{goddemeier2015investigation}.\par

By using the two equations given in (\ref{PDFGClosedForm}) and (\ref{RV}), and using the equation (\ref{SINR}) by omitting its \ac{ICI} variable $E_{x,q}\left|\mathbf{H}(p, q ,f(t_i))\right|^2$, the final form of the random variable \ac{SNR} can be derived as 

\begin{equation}
\begin{aligned}
f_{\gamma_p}(\gamma_p) & =\int_0^{\infty} \frac{E_{x,\omega}\left|\omega(t_i)\right|}{\zeta} f_{\mathbb{G}}\left(\frac{E_{x,\omega}\left|\omega(t_i)\right| \gamma_p}{\zeta}\right) f_\zeta(\zeta) d \zeta,
\end{aligned}
\end{equation}

\begin{equation}
\begin{aligned}
\hspace{-10pt}
&f_{\mathbb{G}}\left(\frac{E_{x,\omega}\left|\omega(t_i)\right| \gamma_p}{\zeta}\right)= \\& 
\frac{A_{{Rx}}\left(\theta^{Rx^{\prime}}, \phi^{Rx}\right) A_{{Tx}}\left(\theta^{Tx^{\prime}}, \phi^{Tx}\right)}{\kappa(N_{Tx-\text{x}}, N_{Tx-\text{y}})} \times \\&
\delta\left(\frac{E_{x,\omega}\left|\omega(t_i)\right| \gamma}{\zeta}-\kappa(N_{Tx-\text{x}}, N_{Tx-\text{y}})\right),
\end{aligned}
\end{equation}

\begin{equation}
\begin{aligned}
f_{\gamma_p}(\gamma_p) =& \frac{A_{{Rx}}\left(\theta^{Rx^{\prime}}, \phi^{Rx}\right) A_{{Tx}}\left(\theta^{Tx^{\prime}}, \phi^{Tx}\right)m^m}{\gamma_p\Gamma(m)}\times \\& 
\int_0^{\infty}\zeta^{m}\exp\left(-m\zeta\right)d\zeta,
\end{aligned}
\end{equation}

\begin{equation}
\begin{aligned}
& f_{\gamma_p}(\gamma_p)  =\frac{A_{{Rx}}\left(\theta^{Rx^{\prime}}, \phi^{Rx}\right) A_{{Tx}}\left(\theta^{Tx^{\prime}}, \phi^{Tx}\right)m^m}{\gamma_p\Gamma(m)}\times \\&
\frac{E_{x,\omega}\left|\omega(t_i)\right|^m \gamma_p^m}{{\kappa(N_{Tx-\text{x}}, N_{Tx-\text{y}})}^m}\exp\left(\frac{-mE_{x,\omega}\left|\omega(t_i)\right|\gamma_p}{\kappa(N_{Tx-\text{x}}, N_{Tx-\text{y}})}\right),
\end{aligned}
\end{equation}

\begin{equation}
\begin{aligned}
\label{PDFSNR}
\hspace{-15pt}
f_{\gamma_p}({\gamma_p}) =&  \frac{\left(E_{x,\omega}\left|\omega(t_i)\right| m\right)^m}{\Gamma(m)}\times\\&
\frac{A_{{Rx}}\left(\theta^{Rx^{\prime}}, \phi^{Rx}\right) A_{{Tx}}\left(\theta^{Tx^{\prime}}, \phi^{Tx}\right)}{\kappa(N_{Tx-\text{x}}, N_{Tx-\text{y}})^m}\gamma_p^{m-1}\times \\& \exp\left(\frac{-m E_{x,\omega}\left|\omega(t_i)\right|\gamma_p}{\kappa(N_{Tx-\text{x}}, N_{Tx-\text{y}})}\right).
\end{aligned}
\end{equation}

Now the effect of the \ac{ICI} can be included in the analysis to find the distribution of the \ac{SINR}. By referring the (\ref{SINR}), and (\ref{PDFSNR}) the \ac{PDF} of the \ac{ICI} for the $q\in\{0, 1, \cdots, N_c-1\}, q\neq p$ can be derived as 

\begin{equation}
\begin{aligned}
\label{PDFICIFinal}
\hspace{-15pt}
f_{\gamma_q}(\gamma_{q}) =& \sum_{q=0}^{Nc-1}\frac{\left(E_{x,\omega}\left|\omega(t_i)\right| m\right)^m}{\Gamma(m)}\times\\&
\frac{A_{{Rx}}\left(\theta^{Rx^{\prime}}, \phi^{Rx}\right) A_{{Tx}}\left(\theta^{Tx^{\prime}}, \phi^{Tx}\right)}{\kappa(N_{Tx-\text{x}}, N_{Tx-\text{y}})^m}\gamma_q^{m-1}\times \\& \exp\left(\frac{-m E_{x,\omega}\left|\omega(t_i)\right|\gamma_q}{\kappa(N_{Tx-\text{x}}, N_{Tx-\text{y}})}\right).
\end{aligned}
\end{equation}

Now, if a random noise due to \ac{ICI}, $\omega_{ICI}$, is appended on the \ac{AWGN}, $\omega(t_i)$, the resulting noise, $\hat{\omega}(t_i)$, can be substituted in (\ref{PDFSNRFinal}) instead of $\omega(t_i)$ to include the effect of \ac{ICI} into the transmitted signal given as 

\begin{equation}
\begin{aligned}
\label{PDFSNRFinal}
\hspace{-15pt}
f_{\hat{\gamma}_p}({\hat{\gamma_p}}) =& \frac{\left(E_{x,\omega}\left|\hat{\omega}(t_i)\right| m\right)^m}{\Gamma(m)}\times\\&
\frac{A_{{Rx}}\left(\theta^{Rx^{\prime}}, \phi^{Rx}\right) A_{{Tx}}\left(\theta^{Tx^{\prime}}, \phi^{Tx}\right)}{\kappa(N_{Tx-\text{x}}, N_{Tx-\text{y}})^m}\hat{\gamma}_p^{m-1}\times \\& \exp\left(\frac{-m E_{x,\omega}\left|\hat{\omega}(t_i)\right|\hat{\gamma_p}}{\kappa(N_{Tx-\text{x}}, N_{Tx-\text{y}})}\right).
\end{aligned}
\end{equation}

After completing this step the final form of the distribution of \ac{SINR} can be given as 

\begin{equation}
    f_{\gamma}(\gamma)=\int_0^{\infty}f_{\hat{\gamma}_p}(\hat{\gamma}_{p})f_{\gamma_q}(\gamma_{q})dp.
\end{equation}

\section{Simulation Results and Discussion} 
\label{Sec:simulation}
In this section, the simulation results of the analyzed channel model is presented in terms of \ac{SINR}, capacity and \ac{PDF} of \ac{SINR}. The parameters used in the simulations are given in Table I.

\begin{figure*}[h]
    \centering
    \subfigure[$N_{Tx}=16$]{{\includegraphics[scale=0.18]{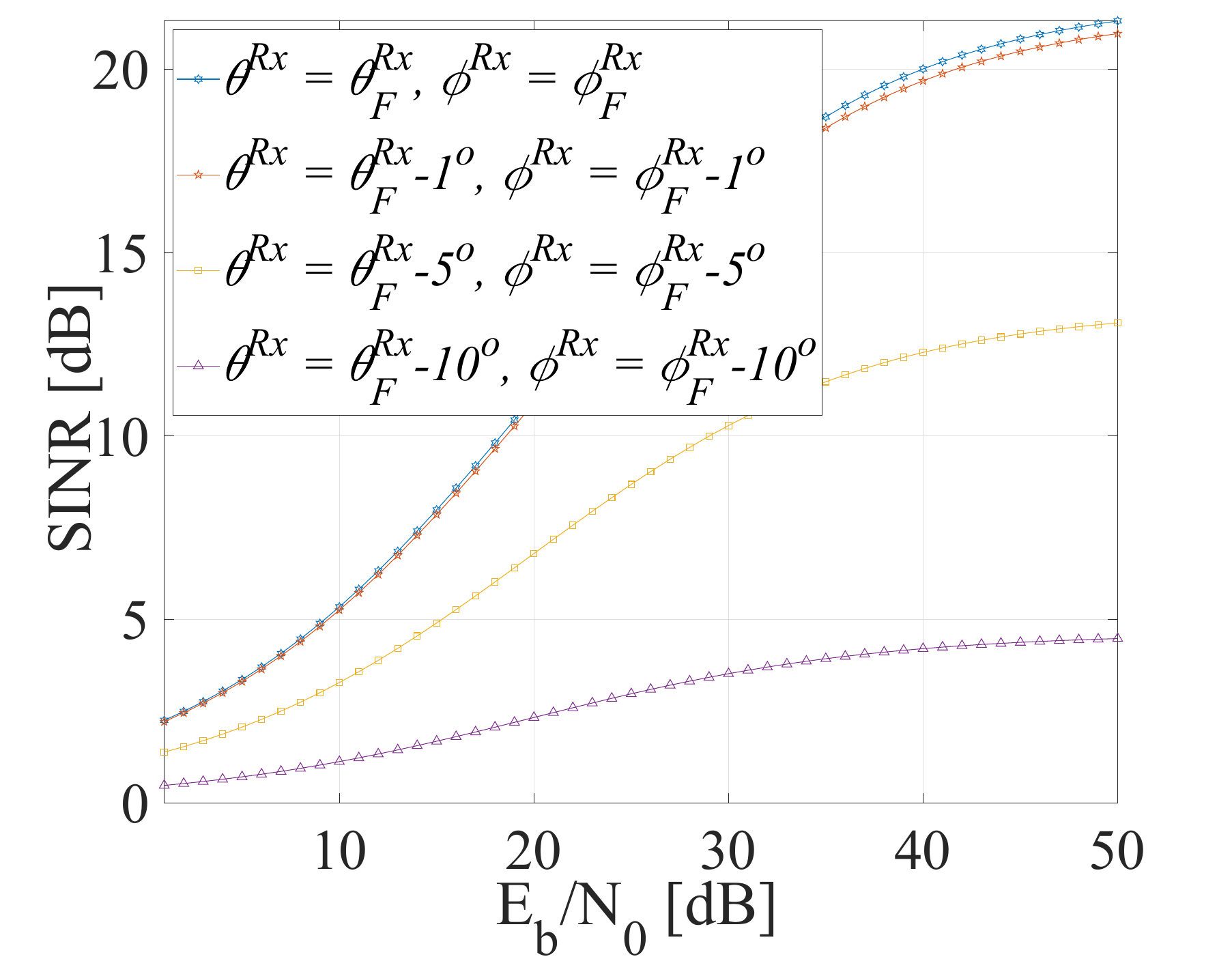}}}%
    \qquad
    \subfigure[$N_{Tx}=32$]{{\includegraphics[scale=0.18]{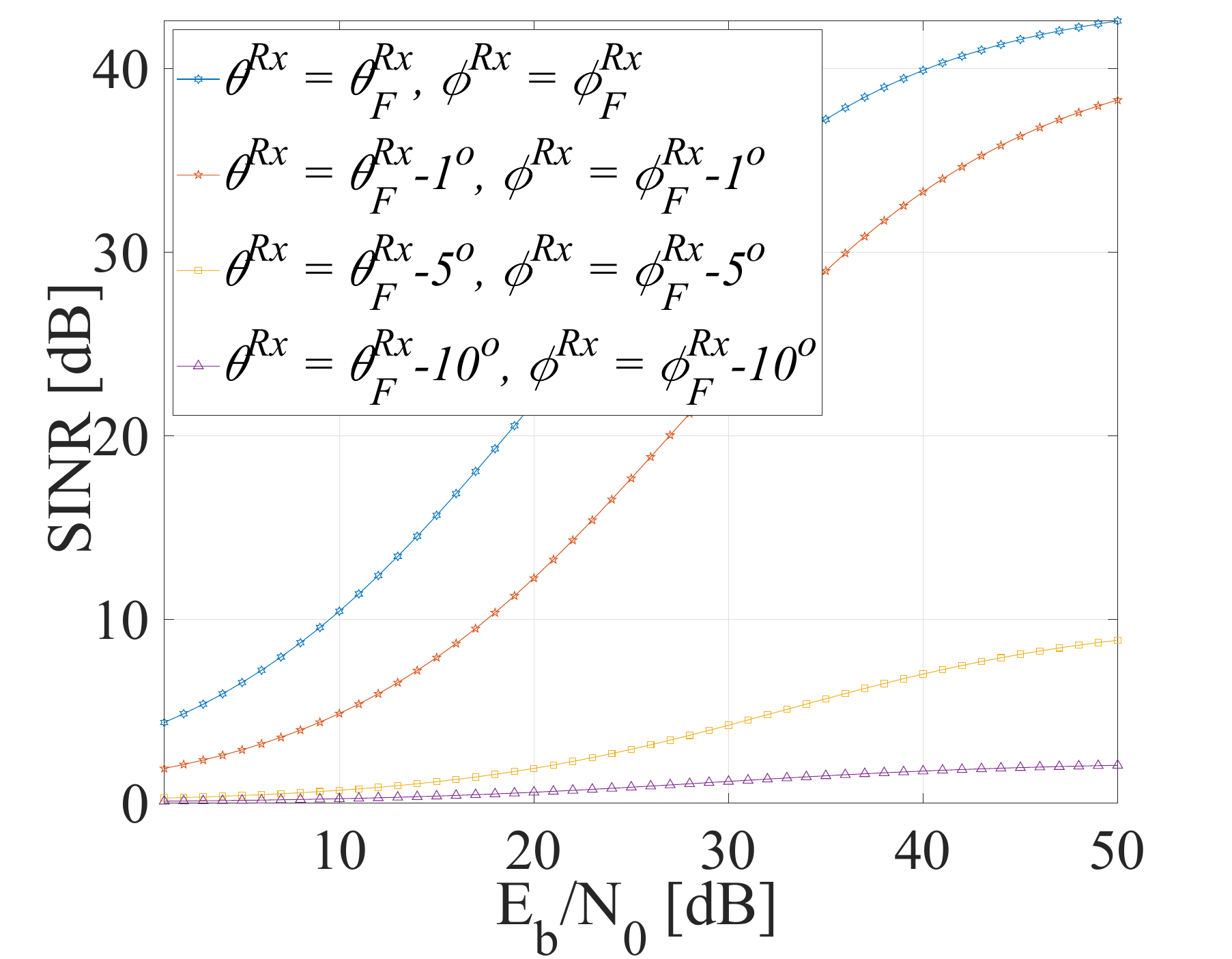}}}%
    \qquad
    \subfigure[$N_{Tx}=64$]{{\includegraphics[scale=0.18]{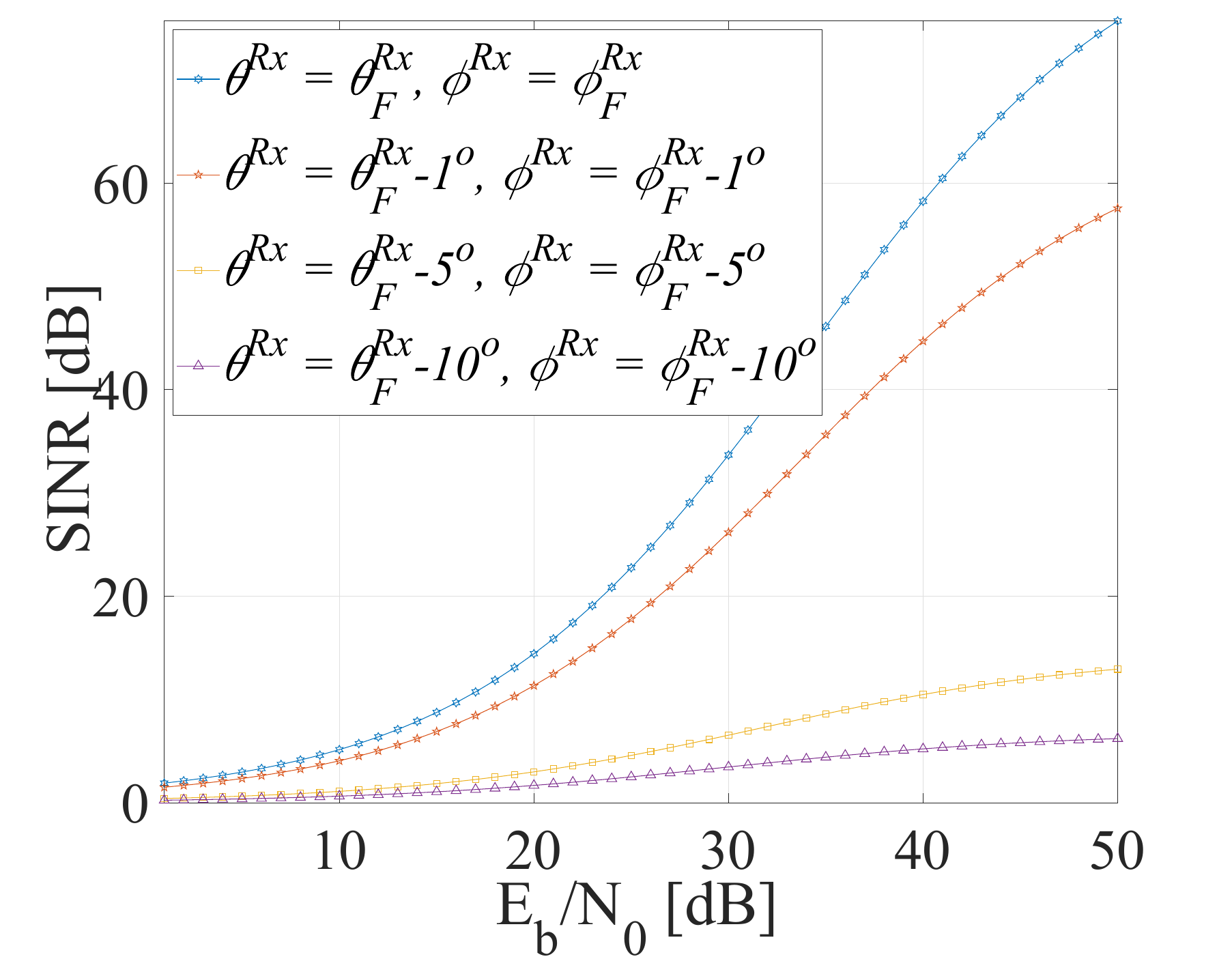}}}%
    \qquad
    \caption{SINR performance of the proposed channel model under different AAOAs and EAOAs when the effect of Doppler is neglected.}%
    \label{SINR1}%
\end{figure*}

\begin{table} [h] 
\begin{center}
\caption{Simulation Parameters \cite{karapantazis2005broadband, akdeniz2014millimeter, ying2020gmd, he2019propagation, li2020reconfigurable}}
\begin{tabular}{|c|c|}
\hline  Altitudes of \ac{HAP}-1 and \ac{HAP}-2 ($\mathbbm{h}_1,\mathbbm{h}_2$)            & $\approx 20$ km               \\
\hline  \ac{3D} distance between \ac{HAP}-1 and \ac{HAP}-2 ($D_{3D}$)                     & $\leq 500$ m                 \\
\hline  Number of transmit antennas on \ac{HAP}-1  ($N_{Tx}$)                             & $16, 32, 64$                      \\
\hline  Number of receive antennas on \ac{HAP}-2 ($N_{Rx}$)                               & $16, 32, 64$                      \\
\hline  Carrier Frequency ($f_c$)                                                         & $60$ GHz                      \\
\hline  Number of Carriers ($N_c$)                                                        & $4$                           \\
\hline  Number of Taps                                                                    & $4$                           \\
\hline  Direction angles in azimuth domain ($\mathcal{D}_{A_{(1,2)}}$)	          & $\left[-\pi, \pi\right]$      \\
\hline  Direction angles in elevation domain ($\mathcal{D}_{E_{(1,2)}}$)         & $\left[-\pi, \pi\right]$      \\
\hline  Phase angles in azimuth domain ($\theta_{1,2} = \theta^{Tx,Rx}$)         & $\left[-\pi, \pi\right]$      \\
\hline  Phase angles in elevation domain ($\phi_{1,2} = \phi^{Tx,Rx}$)           & $\left[-\pi, \pi\right]$      \\
\hline  Focus angles  ($\theta_{F}^{Rx}, \phi_{F}^{Rx}$)  & $\left(60^o,30^o\right)$          \\
\hline
\end{tabular} 
\end{center}
\label{simpar}
\end{table} 
In Fig. \ref{SINR1}, the \ac{SNR} vs. \ac{SINR} performance is exhibited for different \acp{AAOA} and \acp{EAOA} by assuming the \acp{HAP} are stationary. In this regard, the figure shows the \ac{SINR} performance in the case where the \ac{AAOA} and \ac{EAOA} of the received signal match with the receiver's focus angle. It also shows the cases where the \acp{AAOA} and \acp{EAOA} of the received signal mismatch with different angles of deviation. It is obvious from the figure that when the \acp{AAOA} and \acp{EAOA} are matched with the focus angles of the receiver, the performance of \ac{SINR} is maximized due to the fact that the directivity gain become maximum.  However, when the received signal's direction is deviated from the focus angle of the receiver, the performance of the \ac{SINR} gets degraded depending on the severeness of the deviation. \par

\begin{figure*}[h]
    \centering
    \subfigure[$N_{Tx}=16$]{{\includegraphics[scale=0.18]{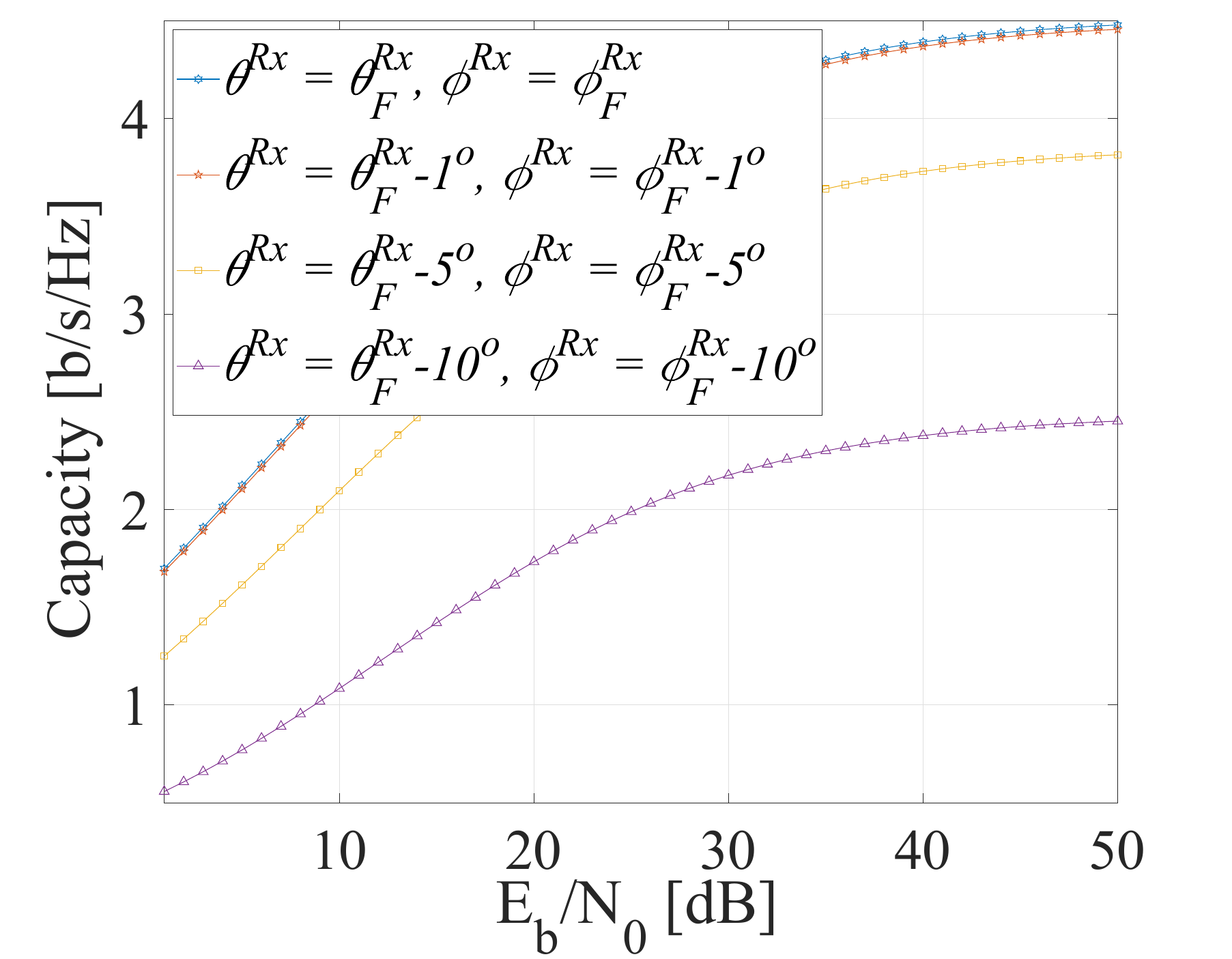}}}%
    \qquad
    \subfigure[$N_{Tx}=32$]{{\includegraphics[scale=0.18]{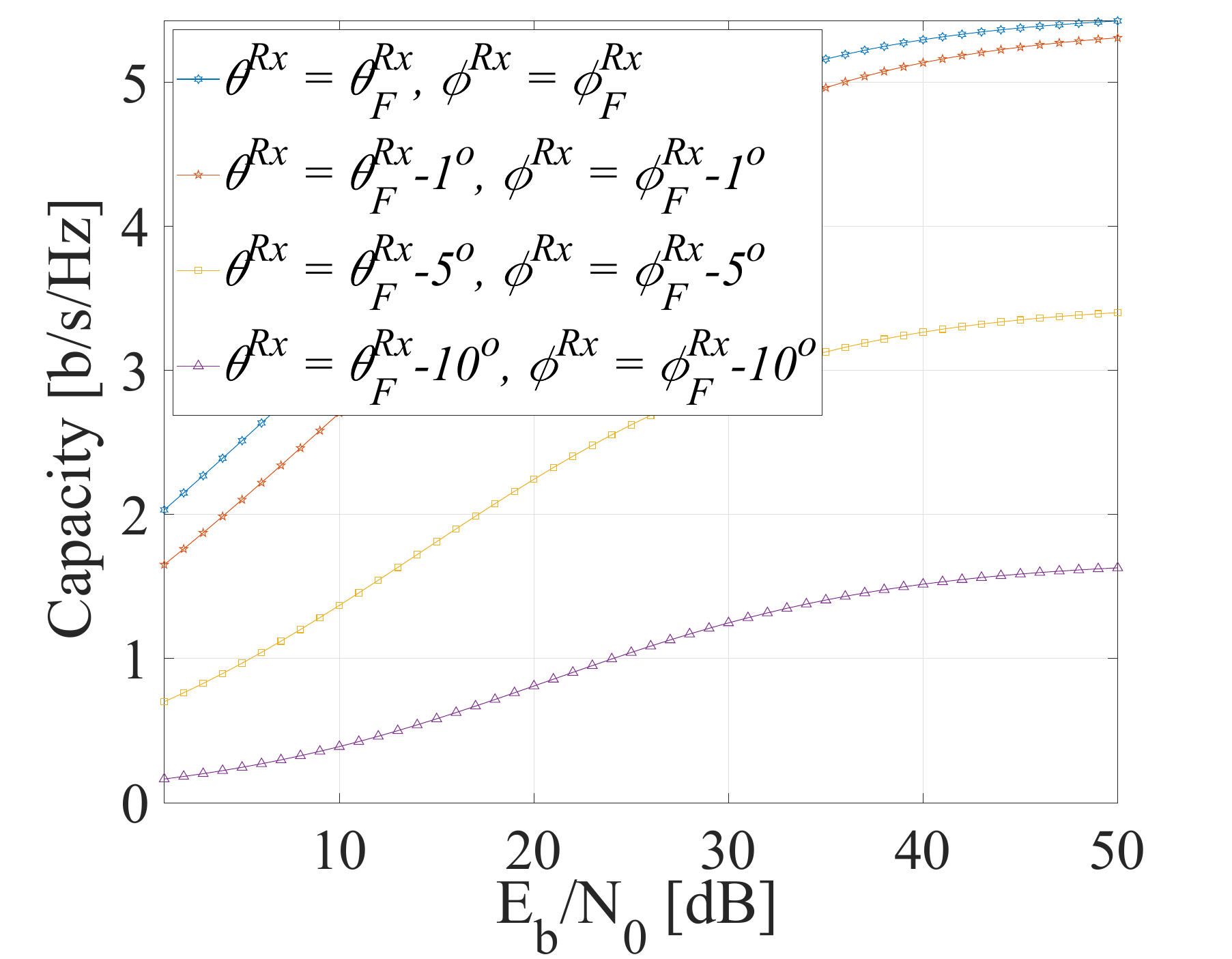}}}%
    \qquad
    \subfigure[$N_{Tx}=64$]{{\includegraphics[scale=0.18]{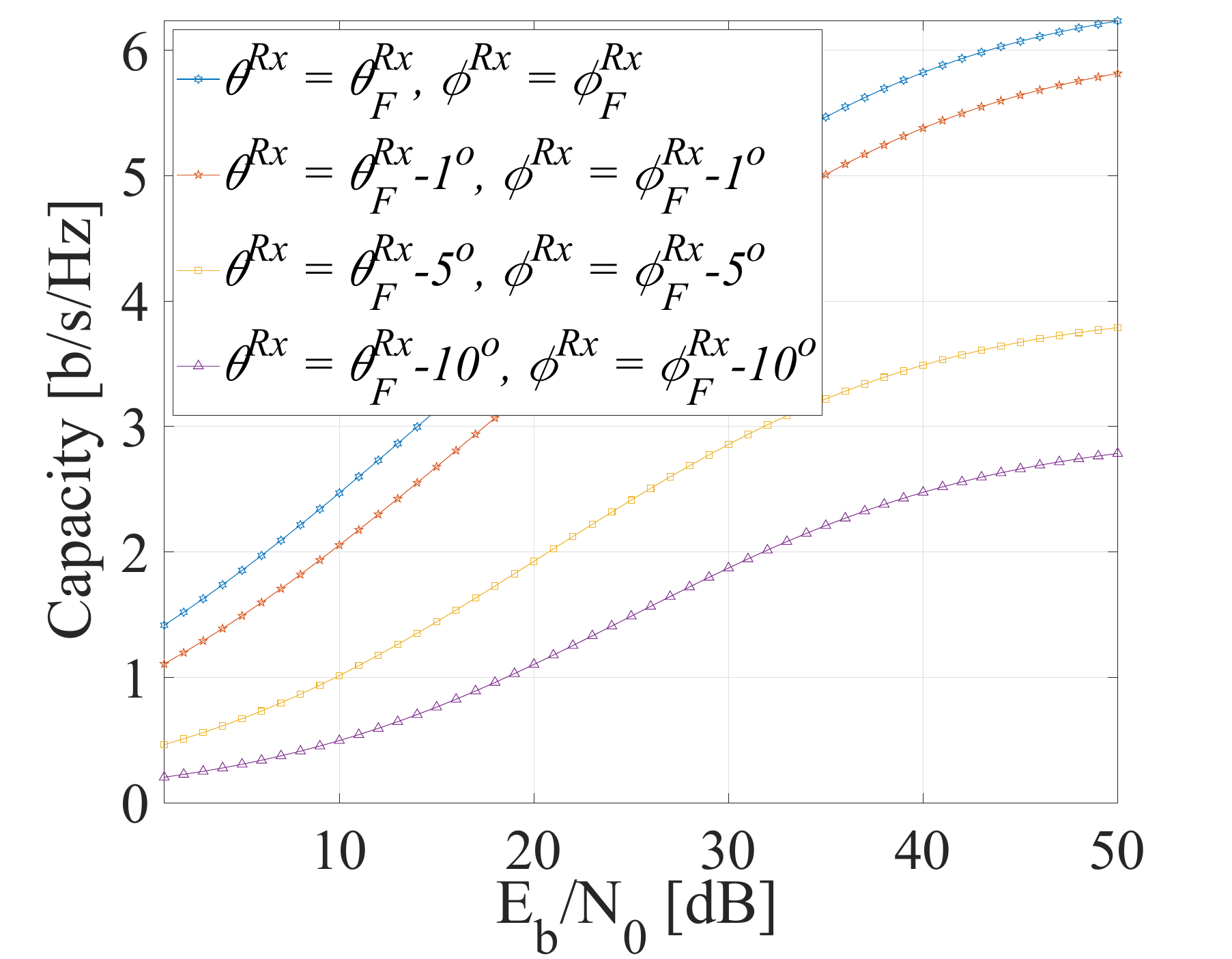}}}%
    \qquad
    \caption{Capacity of the proposed channel model under different AAOAs and EAOAs when the effect of Doppler is neglected.}%
    \label{Capacity1}%
\end{figure*}

In Fig. \ref{Capacity1}, the capacity of the system is shown for different \acp{AAOA} and \acp{EAOA}. As it can be seen from this figure, similar to Fig. \ref{SINR1}, when the \acp{AAOA} and \acp{EAOA} are matched with the focus angles of the receiver, the capacity is maximized, whereas as the \acp{AAOA}' and \acp{EAOA}' deviation increases the capacity decreases accordingly. \par

\begin{figure}[h]
\centering
\includegraphics[scale=0.4]{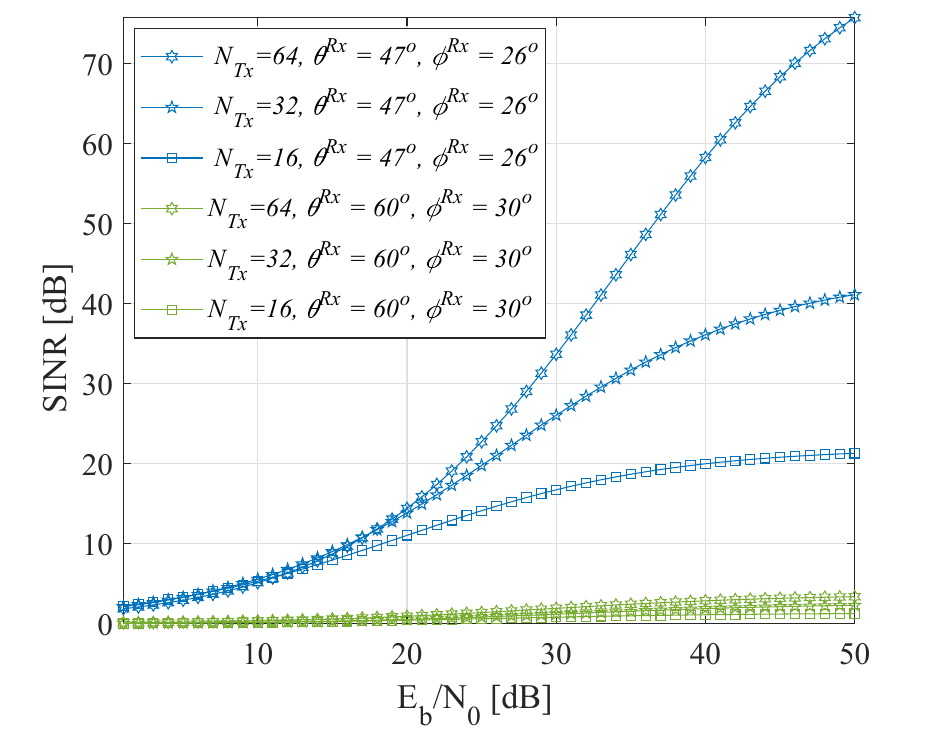}
\caption{\ac{SINR} performance of the proposed channel model in the case where the \acp{HAP} are mobile and the focus angles are defined as $(\theta_{F}^{Rx}, \phi_{F}^{Rx}) = (60^o,30^o)$}
\label{SINR2}
\end{figure}

In Fig. \ref{SINR2}, the \ac{SINR} performance of the proposed channel model is interrogated under the condition where both of the \acp{HAP} are mobile and the focus angle of the receiver is defined to be $(\theta_{F}^{Rx}, \phi_{F}^{Rx}) = (60^o,30^o)$. As shown in Fig. \ref{gainNoDoppler}, when the effect of Doppler is included due to mobility, there is a shift on the focus angles of the receiver, which leads the maximum beam gain to be achieved at a different set of angles rather than the desired one. Depending on the shift on the focus angles of the receiver, it is expected that the \ac{SINR} performance of the system will not be maximized at the desired focus angles. As it can be seen from Fig. \ref{SINR2}, even though the desired focus angles are defined to be $(\theta_{F}^{Rx}, \phi_{F}^{Rx}) = (60^o,30^o)$, due to the Doppler, the maximum \ac{SINR} is achieved at $(\theta^{Rx}, \phi^{Rx}) = (47^o,26^o)$.
\par

\begin{figure}[h]
\centering
\includegraphics[scale=0.40]{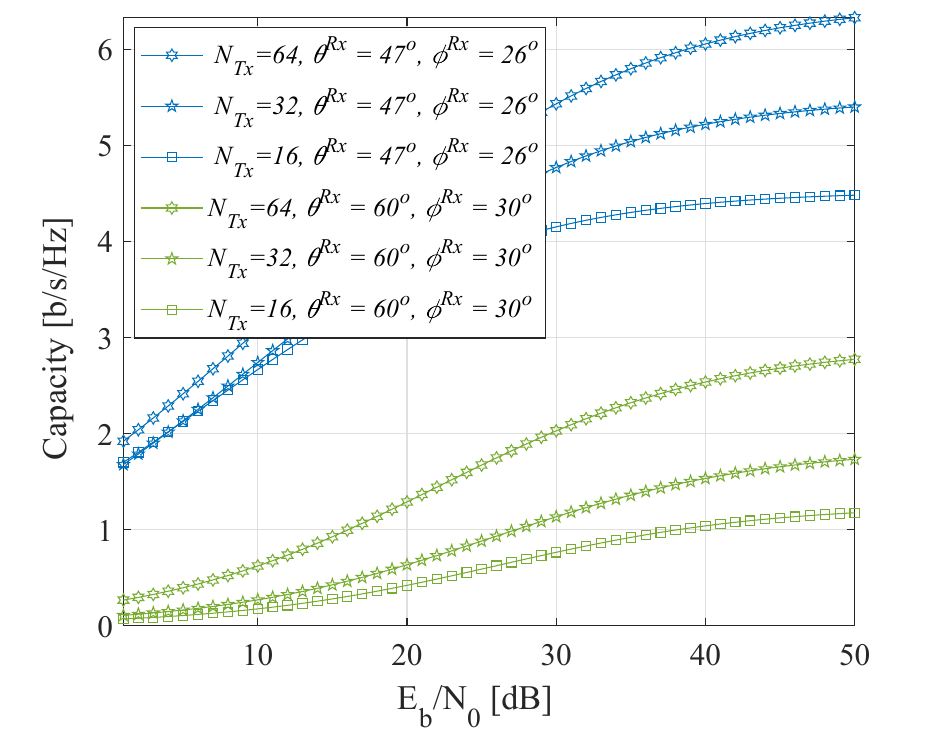}
\caption{Capacity performance of the proposed channel model in the case where the \acp{HAP} are mobile and the focus angles are defined as $(\theta_{F}^{Rx}, \phi_{F}^{Rx}) = (60^o,30^o)$}
\label{Capacity2}
\end{figure}

Fig. \ref{Capacity2} shows the capacity of the system in the case where both of the \acp{HAP} are mobile and the focus angle of the receiver is defined to be $(\theta_{F}^{Rx}, \phi_{F}^{Rx}) = (60^o,30^o)$. As it can be seen from this figure, while the performance of the \ac{SINR} is expected to be higher at the angles $(\theta^{Rx}, \phi^{Rx}) = (60^o,30^o)$, due to the mobility the focus angles are shifted and the \ac{SINR} performance at the angles $(\theta^{Rx}, \phi^{Rx}) = (47^o,26^o)$ outperforms the \ac{SINR} performance at the angles $(\theta^{Rx}, \phi^{Rx}) = (60^o,30^o)$.

\begin{figure}[h]
\centering
\includegraphics[scale=0.40]{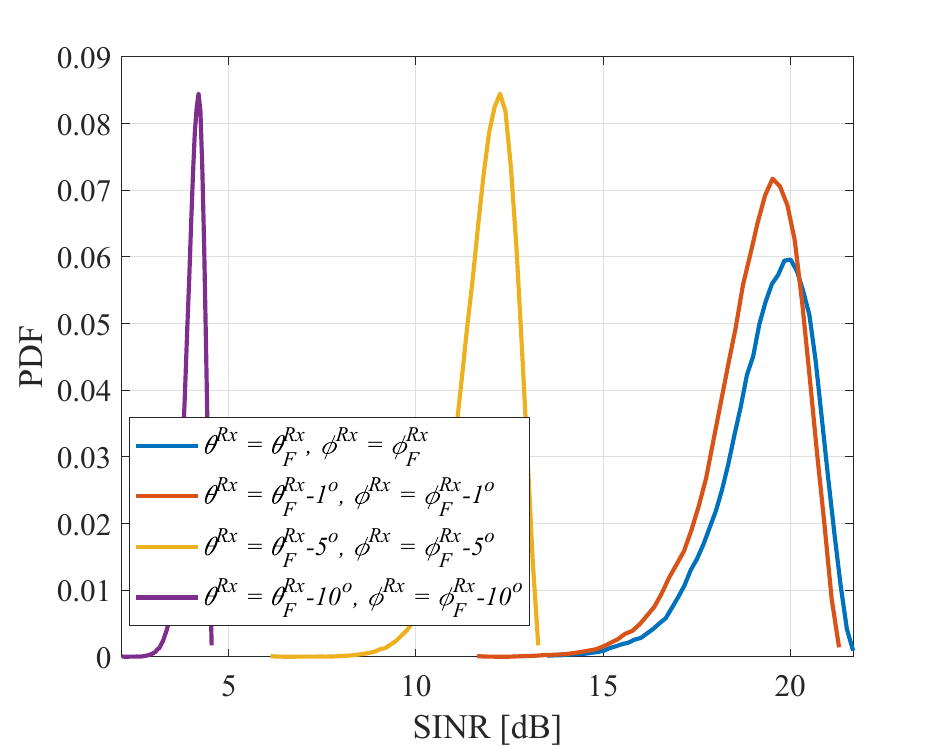}
\caption{\ac{PDF} of the \ac{SINR} under different \acp{AAOA} and \acp{EAOA} deviations when $N_{Tx}=16$ and the effect of Doppler is neglected.}
\label{pdf1}
\end{figure}
\begin{figure}[h]
\centering
\includegraphics[scale=0.40]{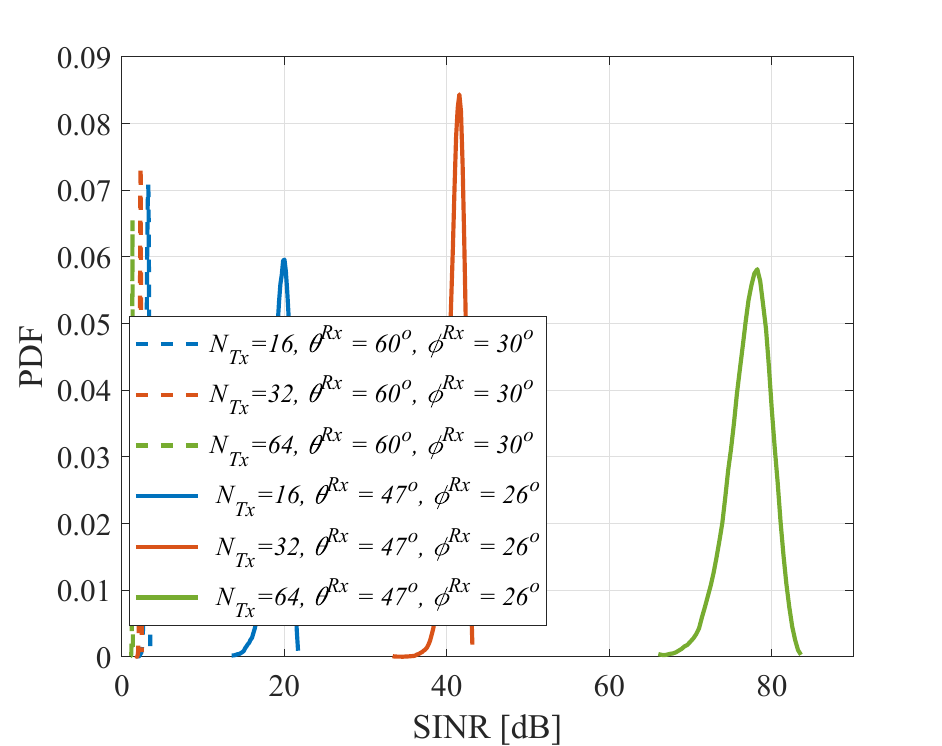}
\caption{\ac{PDF} of the \ac{SINR} in the case where the \acp{HAP} are mobile and the focus angles are defined as $(\theta_{F}^{Rx}, \phi_{F}^{Rx}) = (60^o,30^o)$.}
\label{pdf2}
\end{figure}

Fig. \ref{pdf1} and Fig. \ref{pdf2} show the \ac{SINR} distribution of the proposed channel model for different deviations in the focus angle in those cases where the effect of Doppler due to mobility is omitted and included respectively. As it can be seen from Fig. \ref{pdf1}, when the \acp{HAP} are stationary, the distribution of the \ac{SINR} is gathering around smaller values as the deviations in \ac{AAOA} and \ac{EAOA} increases. In Fig. \ref{pdf2}, the distribution of \ac{SINR} is shown in the case where the \acp{HAP} are mobile. In this figure, even though the maximum \ac{SINR} is expected to be achieved at the desired focus angles, due to the effect of Doppler, it is observed at a different set of angles.\par  
\begin{figure}[h]
\centering
\includegraphics[scale=0.40]{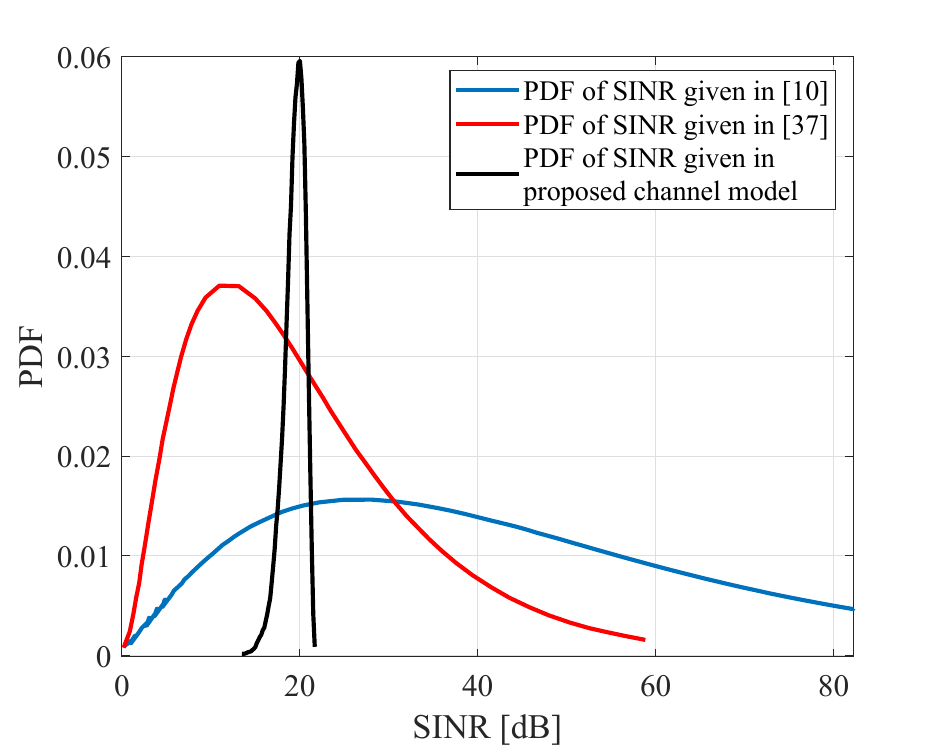}
\caption{\ac{PDF} comparison of the proposed channel model with the related works in the literature.}
\label{pdf3}
\end{figure}
In Fig. \ref{pdf3}, a comparison between the proposed channel model and two other related works in the literature is given to emphasize the novelty of our work. The main difference that is desired to be shown in this figure is the effect of mobility. The \ac{PDF} of \cite{dabiri2020analytical} shown in blue line is given for a low altitude \ac{UAV} scenario where the \acp{UAV} are interrogated under hovering conditions. Moreover, the \ac{PDF} of  \cite{ozdemir2008opportunistic} illustrated in red line is given for a case where the opportunistic beamforming is performed in a multi-user scenario for a scheduled user with partial channel information. Even though the results given in \cite{ozdemir2008opportunistic} do not provide an analysis for \ac{A2A} channel, it is still a proper comparison material to provide an insight for the evaluation of the difference between the stationary and mobile scenarios in beamforming applications. As it can be seen from the overall of this figure, the \ac{PDF} of the \ac{SINR} is lower at the two other scenarios because of the fact that the effect of \ac{ICI} due to the mobility is not considered. However, when the effect of mobility is taken into account and compensated, the distribution of \ac{SINR} increases accordingly. 
\section {Conclusion} 
\label{Sec:conclusion}
In this paper, an analytical \ac{3D} channel model operating at \ac{mmWave} band is proposed to represent the beamforming characteristics of \acp{HAP} in future \ac{6G} \acp{FANET}. In this regard, the \ac{MIMO} characteristics of the two \acp{HAP} that are considered to be equipped with $N\times N$ \ac{URPA} antennas are interrogated in the presence of high mobility and its resultant Doppler shift. Moreover, a joint Gauss-Markov/\ac{3D} random walk mobility model is introduced to provide a more accurate movement estimation of \acp{HAP} in real-life scenarios. The suggested model in this paper is proven by mathematical derivations and their validity is consolidated by Monte-Carlo simulations. The resulting outcomes from the obtained calculations and simulations in terms of \ac{SINR}, capacity, gain, and \ac{PDF} prove that the applicability of such communication that exploits beamforming in high frequencies between two mobile \acp{HAP} has a great potential to satisfy the requirements of future airborne \acp{NTN} may present.   

%\section*{Acknowledgment}
%This work was supported by the Scientific and Technological Research Council of %Turkey (TÜB\.{I}TAK) under Grant XXXXXXX, with the cooperation of Istanbul %Medipol University. 

% \ifCLASSOPTIONcaptionsoff
% %  \newpage
% \fi

%\bibliography{IEEEfull,references}

\begin{thebibliography}{10}
\providecommand{\url}[1]{#1}
\csname url@samestyle\endcsname
\providecommand{\newblock}{\relax}
\providecommand{\bibinfo}[2]{#2}
\providecommand{\BIBentrySTDinterwordspacing}{\spaceskip=0pt\relax}
\providecommand{\BIBentryALTinterwordstretchfactor}{4}
\providecommand{\BIBentryALTinterwordspacing}{\spaceskip=\fontdimen2\font plus
\BIBentryALTinterwordstretchfactor\fontdimen3\font minus \fontdimen4\font\relax}
\providecommand{\BIBforeignlanguage}[2]{{%
\expandafter\ifx\csname l@#1\endcsname\relax
\typeout{** WARNING: IEEEtran.bst: No hyphenation pattern has been}%
\typeout{** loaded for the language `#1'. Using the pattern for}%
\typeout{** the default language instead.}%
\else
\language=\csname l@#1\endcsname
\fi
#2}}
\providecommand{\BIBdecl}{\relax}
\BIBdecl

\bibitem{li2018uav}
B.~Li, Z.~Fei, and Y.~Zhang, ``{UAV} communications for 5{G} and beyond: Recent advances and future trends,'' \emph{IEEE Internet of Things Journal}, vol.~6, no.~2, pp. 2241--2263, 2018.

\bibitem{kirik2023inter}
M.~K{\i}r{\i}k, N.~A. Abusanad, and H.~Arslan, ``Inter-{HAP} based geometrical 3-{D} channel model operating at 28 to 60 {GH}z for future 6{G} non-terrestrial networks,'' in \emph{IEEE Wireless Communications and Networking Conference (WCNC)}.\hskip 1em plus 0.5em minus 0.4em\relax IEEE, 2023, pp. 1--5.

\bibitem{demir2022performance}
Y.~I. Demir, M.~S.~J. Solaija, and H.~Arslan, ``On the performance of handover mechanisms for non-terrestrial networks,'' in \emph{IEEE 95th Vehicular Technology Conference:(VTC2022-Spring)}.\hskip 1em plus 0.5em minus 0.4em\relax IEEE, 2022, pp. 1--5.

\bibitem{zhang20196g}
Z.~Zhang, Y.~Xiao, Z.~Ma, M.~Xiao, Z.~Ding, X.~Lei, G.~K. Karagiannidis, and P.~Fan, ``6{G} wireless networks: Vision, requirements, architecture, and key technologies,'' \emph{IEEE Vehicular Technology Magazine}, vol.~14, no.~3, pp. 28--41, 2019.

\bibitem{afeef2022beam}
L.~Afeef and H.~Arslan, ``Beam squint effect in multi-beam mm{W}ave massive {MIMO} systems,'' in \emph{IEEE 96th Vehicular Technology Conference (VTC2022-Fall)}.\hskip 1em plus 0.5em minus 0.4em\relax IEEE, 2022, pp. 1--5.

\bibitem{wang2018millimeter}
X.~Wang, L.~Kong, F.~Kong, F.~Qiu, M.~Xia, S.~Arnon, and G.~Chen, ``Millimeter wave communication: A comprehensive survey,'' \emph{IEEE Communications Surveys \& Tutorials}, vol.~20, no.~3, pp. 1616--1653, 2018.

\bibitem{ma2019wideband}
Z.~Ma, B.~Ai, R.~He, G.~Wang, Y.~Niu, and Z.~Zhong, ``A wideband non-stationary air-to-air channel model for {UAV} communications,'' \emph{IEEE Transactions on Vehicular Technology}, vol.~69, no.~2, pp. 1214--1226, 2019.

\bibitem{ma2020three}
Z.~Ma, B.~Ai, R.~He, Z.~Zhong, M.~Yang, J.~Wang, L.~Pei, Y.~Li, and J.~Li, ``Three-dimensional modeling of millimeter-wave {MIMO} channels for {UAV}-based communications,'' in \emph{GLOBECOM}.\hskip 1em plus 0.5em minus 0.4em\relax IEEE, 2020, pp. 1--6.

\bibitem{ma2021non}
Z.~Ma, B.~Ai, R.~He, Z.~Zhong, and M.~Yang, ``A non-stationary geometry-based {MIMO} channel model for millimeter-wave {UAV} networks,'' \emph{IEEE Journal on Selected Areas in Communications}, vol.~39, no.~10, pp. 2960--2974, 2021.

\bibitem{dabiri2020analytical}
M.~T. Dabiri, H.~Safi, S.~Parsaeefard, and W.~Saad, ``Analytical channel models for millimeter wave {UAV} networks under hovering fluctuations,'' \emph{IEEE Transactions on Wireless Communications}, vol.~19, no.~4, pp. 2868--2883, 2020.

\bibitem{jiang2019three}
H.~Jiang, Z.~Zhang, and G.~Gui, ``Three-dimensional non-stationary wideband geometry-based {UAV} channel model for {A2G} communication environments,'' \emph{IEEE Access}, vol.~7, pp. 26\,116--26\,122, 2019.

\bibitem{zeng20173d}
L.~Zeng, X.~Cheng, C.-X. Wang, and X.~Yin, ``A 3{D} geometry-based stochastic channel model for {UAV}-{MIMO} channels,'' in \emph{IEEE Wireless Communications and Networking Conference (WCNC)}.\hskip 1em plus 0.5em minus 0.4em\relax IEEE, 2017, pp. 1--5.

\bibitem{mao20213d}
X.~Mao, C.-X. Wang, and H.~Chang, ``A 3{D} non-stationary geometry-based stochastic model for 6{G} {UAV} air-to-air channels,'' in \emph{13th International Conference on Wireless Communications and Signal Processing (WCSP)}.\hskip 1em plus 0.5em minus 0.4em\relax IEEE, 2021, pp. 1--5.

\bibitem{jiang2023physics}
H.~Jiang, B.~Xiong, H.~Zhang, and E.~Basar, ``Physics-based 3{D} end-to-end modeling for double-{RIS} assisted non-stationary {UAV}-to-ground communication channels,'' \emph{IEEE Transactions on Communications}, 2023.

\bibitem{cao2021non}
C.~Cao, Z.~Lian, Y.~Wang, Y.~Su, and B.~Jin, ``A non-stationary geometry-based channel model for {IRS}-assisted {UAV}-{MIMO} channels,'' \emph{IEEE Communications Letters}, vol.~25, no.~12, pp. 3760--3764, 2021.

\bibitem{lian2021non}
Z.~Lian, Y.~Su, Y.~Wang, and L.~Jiang, ``A non-stationary 3-{D} wideband channel model for intelligent reflecting surface-assisted {HAP}-{MIMO} communication systems,'' \emph{IEEE Transactions on Vehicular Technology}, vol.~71, no.~2, pp. 1109--1123, 2021.

\bibitem{lian2016novel}
Z.~Lian, L.~Jiang, C.~He, and Q.~Xi, ``A novel channel model for 3-{D} {HAP}-{MIMO} communication systems,'' in \emph{International Conference on Networking and Network Applications (NaNA)}.\hskip 1em plus 0.5em minus 0.4em\relax IEEE, 2016, pp. 1--6.

\bibitem{lian2018non}
Z.~Lian, L.~Jiang, C.~He, and D.~He, ``A non-stationary 3-{D} wideband {GBSM} for {HAP}-{MIMO} communication systems,'' \emph{IEEE Transactions on Vehicular Technology}, vol.~68, no.~2, pp. 1128--1139, 2018.

\bibitem{liang1999predictive}
B.~Liang and Z.~J. Haas, ``Predictive distance-based mobility management for pcs networks,'' in \emph{IEEE INFOCOM'99. Conference on Computer Communications. Proceedings. Eighteenth Annual Joint Conference of the IEEE Computer and Communications Societies. The Future is Now (Cat. No. 99CH36320)}, vol.~3.\hskip 1em plus 0.5em minus 0.4em\relax IEEE, 1999, pp. 1377--1384.

\bibitem{chiang20042}
K.-H. Chiang and N.~Shenoy, ``A 2-{D} random-walk mobility model for location-management studies in wireless networks,'' \emph{IEEE Transactions on Vehicular Technology}, vol.~53, no.~2, pp. 413--424, 2004.

\bibitem{camp2002survey}
T.~Camp, J.~Boleng, and V.~Davies, ``A survey of mobility models for ad hoc network research,'' \emph{Wireless Communications and Mobile Computing}, vol.~2, no.~5, pp. 483--502, 2002.

\bibitem{ariyakhajorn2006comparative}
J.~Ariyakhajorn, P.~Wannawilai, and C.~Sathitwiriyawong, ``A comparative study of random waypoint and {G}auss-{M}arkov mobility models in the performance evaluation of manet,'' in \emph{International Symposium on Communications and Information Technologies}.\hskip 1em plus 0.5em minus 0.4em\relax IEEE, 2006, pp. 894--899.

\bibitem{bekmezci2013flying}
I.~Bekmezci, O.~K. Sahingoz, and {\c{S}}.~Temel, ``Flying ad-hoc networks (fanets): A survey,'' \emph{Ad Hoc Networks}, vol.~11, no.~3, pp. 1254--1270, 2013.

\bibitem{aguiar2008solid}
P.~Aguiar, D.~Brett, and N.~Brandon, ``Solid oxide fuel cell/gas turbine hybrid system analysis for high-altitude long-endurance unmanned aerial vehicles,'' \emph{International Journal of Hydrogen Energy}, vol.~33, no.~23, pp. 7214--7223, 2008.

\bibitem{ma20193d}
Z.~Ma, B.~Ai, R.~He, and Z.~Zhong, ``A 3{D} air-to-air wideband non-stationary channel model of {UAV} communications,'' in \emph{90th Vehicular Technology Conference (VTC2019-Fall)}.\hskip 1em plus 0.5em minus 0.4em\relax IEEE, 2019, pp. 1--5.

\bibitem{tan2017analysis}
W.~Tan, S.~D. Assimonis, M.~Matthaiou, Y.~Han, X.~Li, and S.~Jin, ``Analysis of different planar antenna arrays for mm{W}ave massive {MIMO} systems,'' in \emph{85th Vehicular Technology Conference (VTC Spring)}.\hskip 1em plus 0.5em minus 0.4em\relax IEEE, 2017, pp. 1--5.

\bibitem{tse2005fundamentals}
D.~Tse and P.~Viswanath, \emph{Fundamentals of wireless communication}.\hskip 1em plus 0.5em minus 0.4em\relax Cambridge university press, 2005.

\bibitem{el2014spatially}
O.~El~Ayach, S.~Rajagopal, S.~Abu-Surra, Z.~Pi, and R.~W. Heath, ``Spatially sparse precoding in millimeter wave {MIMO} systems,'' \emph{IEEE Transactions on Wireless Communications}, vol.~13, no.~3, pp. 1499--1513, 2014.

\bibitem{umeyama1988eigendecomposition}
S.~Umeyama, ``An eigendecomposition approach to weighted graph matching problems,'' \emph{IEEE Transactions on Pattern Analysis and Machine Intelligence}, vol.~10, no.~5, pp. 695--703, 1988.

\bibitem{yu2017coverage}
X.~Yu, J.~Zhang, M.~Haenggi, and K.~B. Letaief, ``Coverage analysis for millimeter wave networks: The impact of directional antenna arrays,'' \emph{IEEE Journal on Selected Areas in Communications}, vol.~35, no.~7, pp. 1498--1512, 2017.

\bibitem{goddemeier2015investigation}
N.~Goddemeier and C.~Wietfeld, ``Investigation of air-to-air channel characteristics and a {UAV} specific extension to the rice model,'' in \emph{2015 IEEE Globecom Workshops (GC Wkshps)}.\hskip 1em plus 0.5em minus 0.4em\relax IEEE, 2015, pp. 1--5.

\bibitem{karapantazis2005broadband}
S.~Karapantazis and F.~Pavlidou, ``Broadband communications via high-altitude platforms: A survey,'' \emph{IEEE Communications Surveys \& Tutorials}, vol.~7, no.~1, pp. 2--31, 2005.

\bibitem{akdeniz2014millimeter}
M.~R. Akdeniz, Y.~Liu, M.~K. Samimi, S.~Sun, S.~Rangan, T.~S. Rappaport, and E.~Erkip, ``Millimeter wave channel modeling and cellular capacity evaluation,'' \emph{IEEE Journal on Selected Areas in Communications}, vol.~32, no.~6, pp. 1164--1179, 2014.

\bibitem{ying2020gmd}
K.~Ying, Z.~Gao, S.~Lyu, Y.~Wu, H.~Wang, and M.-S. Alouini, ``{GMD}-based hybrid beamforming for large reconfigurable intelligent surface assisted millimeter-wave massive {MIMO},'' \emph{IEEE Access}, vol.~8, pp. 19\,530--19\,539, 2020.

\bibitem{he2019propagation}
R.~He, C.~Schneider, B.~Ai, G.~Wang, Z.~Zhong, D.~A. Dupleich, R.~S. Thomae, M.~Boban, J.~Luo, and Y.~Zhang, ``Propagation channels of 5{G} millimeter-wave vehicle-to-vehicle communications: Recent advances and future challenges,'' \emph{IEEE Vehicular Technology Magazine}, vol.~15, no.~1, pp. 16--26, 2019.

\bibitem{li2020reconfigurable}
S.~Li, B.~Duo, X.~Yuan, Y.-C. Liang, and M.~Di~Renzo, ``Reconfigurable intelligent surface assisted {UAV} communication: Joint trajectory design and passive beamforming,'' \emph{IEEE Wireless Communications Letters}, vol.~9, no.~5, pp. 716--720, 2020.

\bibitem{ozdemir2008opportunistic}
O.~Ozdemir and M.~Torlak, ``Opportunistic beamforming over rayleigh channels with partial side information,'' \emph{IEEE Transactions on Wireless Communications}, vol.~7, no.~9, pp. 3417--3427, 2008.

\end{thebibliography}
%\bibliographystyle{IEEEtran}

%\input{biography}

% Generated by IEEEtran.bst, version: 1.14 (2015/08/26)

\end{document}